\documentclass[aps,twocolumn,superscriptaddress]{revtex4-2}
\usepackage{graphicx,color,setspace}

\usepackage[version=3]{mhchem} 
\usepackage[permil]{overpic}
\usepackage{tikz}
\usepackage[hidelinks]{hyperref}
\usepackage{graphicx,color,setspace}
\usepackage[utf8]{inputenc}
\usepackage{mathtools}
\usepackage[normalem]{ulem}
\usepackage{tabularx}
\usepackage{multirow}
\usepackage{float}
\usepackage{amsmath}
\usepackage{xr}

\begin{document}

\author{Matthew Sample}
\affiliation{School for Engineering of Matter, Transport, and Energy, Arizona State University, Tempe, AZ 85287, USA}
\affiliation{School of Molecular Sciences and Center for Molecular Design and Biomimetics, The Biodesign Institute, Arizona State University, 1001 South McAllister Avenue, Tempe, Arizona 85281, USA}
\author{Hao Liu}
\affiliation{School of Molecular Sciences and Center for Molecular Design and Biomimetics, The Biodesign Institute, Arizona State University, 1001 South McAllister Avenue, Tempe, Arizona 85281, USA}
\author{Thong Diep}
\affiliation{School of Molecular Sciences and Center for Molecular Design and Biomimetics, The Biodesign Institute, Arizona State University, 1001 South McAllister Avenue, Tempe, Arizona 85281, USA}
\author{Michael Matthies}
\affiliation{School of Molecular Sciences and Center for Molecular Design and Biomimetics, The Biodesign Institute, Arizona State University, 1001 South McAllister Avenue, Tempe, Arizona 85281, USA}
\affiliation{TU Munich, School of Natural Sciences, Department of Bioscience, Garching, Germany}
\author{Petr \v{S}ulc}
\affiliation{School of Molecular Sciences and Center for Molecular Design and Biomimetics, The Biodesign Institute, Arizona State University, 1001 South McAllister Avenue, Tempe, Arizona 85281, USA}
\affiliation{TU Munich, School of Natural Sciences, Department of Bioscience, Garching, Germany}
\title[Hairygami: Analysis of DNA Nanostructures' Conformational Change Driven by Functionalizable Overhangs]{Hairygami: Analysis of DNA Nanostructure's Conformational Change Driven by Functionalizable Overhangs}

\begin{abstract}
DNA origami is a widely used method to construct nanostructures by self-assembling designed DNA strands. These structures are often used as "pegboards" for templated assembly of proteins, gold nanoparticles, aptamers, and other molecules, with applications ranging from therapeutics and diagnostics to plasmonics and photonics.
 Imaging these structures using AFM or TEM does not capture their full conformation ensemble as they only show their shape flattened on a surface. However, certain conformations of the nanostructure can position guest molecules into distances unaccounted for in their intended design, thus leading to spurious interactions between guest molecules that are designed to be separated.
Here, we use molecular dynamics simulations to capture conformational ensemble of 2D DNA origami tiles and show that introducing single-stranded overhangs, which are typically used for functionalization of the origami with guest molecules, induces a curvature of the tile structure in the bulk.  We show that the shape deformation is of entropic origin, with implications for design of robust DNA origami breadboards as well as potential approach to modulate structure shape by introducing overhangs. We then verify experimentally that the DNA overhangs introduce curvature into the DNA origami tiles in divalent as well as monovalent salt buffer conditions. We further experimentally verify that DNA origami functionalized with attached proteins also experience such induced curvature. We provide the developed simulation code implementing the enhanced sampling to characterize conformational space of DNA origami as open source software. 
\end{abstract}

\maketitle
\section{Introduction}

  The emerging fields of DNA and RNA nanotechnology use DNA or RNA strands to self-assemble nanoscale structures and devices. The fields have multiple promising applications that include therapeutics, diagnostics, molecular computing, biotemplatemplated assembly for nanophotonics and optical computing, nano-electronics, and synthetic biology \cite{seeman2007overview,hong2020precise,guo2010emerging,pinheiro2011challenges,kim2020rna}.
  Currently, the most popular construct in DNA nanotechnology is the DNA origami \cite{rothemund2006folding}. It typically consists of a single-stranded DNA scaffold strand taken from M$13$ bacteriophage ($7249$ bases long), and short staple strands that are complementary to different regions of the scaffold strand that then self-assemble a structure of a desired shape. Originally, DNA origami were designed as 2D structures, and later work has extended this concept to $3$D \cite{douglas2009self}. The origami designs have been very quickly adopted by the broad bionanotechnology research community because DNA strands can be functionalized e.g. by attaching gold nanoparticles, proteins, small molecules, quantum dots and aptamers \cite{dey2021dna,ding2010gold,pan2019aptamer,sacca2010orthogonal}. Given the fact that we know where each nucleotide is going to be positioned with respect to the rest of the structure in the final assembled shape, the DNA origami technique effectively provides nanoscale precision for positioning objects with respect to each other.
  A common strategy to introduce these functional moieties to DNA origami is to extend the strands comprising the structure with single-stranded overhangs. However, it has not been previously explored how such modifications can affect the structure of the origami, and therefore potentially its function as well.





    Typically, DNA origami structures are characterized by surface-based techniques like atomic force microscopy (AFM) and transmission electron microscopy (TEM). During this analysis the structure is adhered to a charged surface, limiting the number of conformations that can be observed. While this is not playing a role for applications in molecular electronics, molecular medicine and diagnostics assays often rely on the interactions of the structures in solution, and hence the images produced by AFM and TEM imaging techniques are not necessarily representative of the conformations that the structures sample in solution. $3$D nanostructures have also been characterized by cryo-EM techniques \cite{bai2012cryo,veneziano2016designer}. However, the image processing and reconstruction of $3$D DNA nanostructures in high resolution remains a challenging process, which relies on automated construction of an ensemble average, and hence it can still miss some conformations.    
    Flexible DNA structures sample in the bulk highly deformed conformations \cite{ni2022direct}, potentially impacting their function such as when different functionalized regions that are not supposed to interact with each other might appear in close proximity, or when the attached particles are not at the distances intended in the design. 

Currently, there is no experimental technique available that would allow for easy, reliable and high-precision characterization of the conformational ensemble of DNA origami nanostructures.  Super-resolution imaging-based approaches, such as DNA-PAINT \cite{schnitzbauer2017super}, or small angle X-ray scattering (SAXS) can provide information about conformations in the bulk \cite{baker2018dimensions}, but ideally need to be accompanied by a model that can inform the measurements for improvement of signal to noise ratio in analysis. Here, we show how computational modeling can provide crucial insight into the flexibility and motion of $2$D DNA origami structures, and in particular focus on conformational ensemble changes induced by DNA overhangs attached to the nanostructure. Modeling of DNA origami can present significant challenges, however, given the sizes of the DNA origami of over $14\,000$ nucleotides. Atomistic resolution modeling is limited to at most microseconds timescales, and hence over the past several years coarse-grained models \cite{ouldridge2011structural,snodin2015introducing,sulc2012sequence,maffeo2020mrdna} and finite-element based predictions approaches have been developed to computationally sample DNA origami mean shape  \cite{lee2022predicting,kim2012quantitative,snodin2019coarse,wang2022planar,deluca2023prediction}. In this study, we use the nucleotide-level coarse-grained model oxDNA \cite{ouldridge2011structural,snodin2015introducing,rovigatti2015comparison,poppleton2023oxdna}, as it has been shown to accurately capture both single-stranded and double-stranded DNA biophysics \cite{doye2013coarse,sengar2021primer}.  It is parameterized to reproduce thermodynamic, mechanical and structural properties of both single-stranded and double-stranded DNA \cite{ouldridge2011structural}. 
The oxDNA model has been previously shown to successfully reproduce a range of DNA properties  \cite{doye2013coarse}, including the radius of gyration of ssDNA at different salt concentrations \cite{snodin2015introducing}, the thermodynamic and kinetic effects of single-stranded stacking in hairpin loops \cite{vsulc2014nucleotide,mosayebi2014role}, as well as the conformations of 2D and 3D DNA origami structures that are consistent with experimental SAXS and cryo-EM measurements \cite{snodin2019coarse}. 

  \begin{figure}
     \centering
     \includegraphics[width=0.5\textwidth]{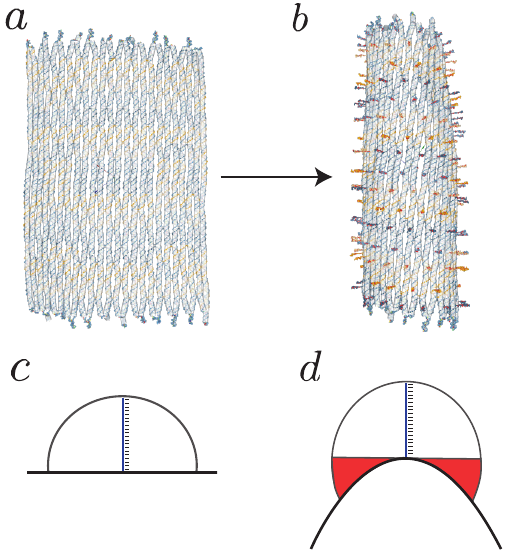}
    \caption{The addition of overhangs causes a $2$D DNA origami tile $(a)$ to adopt a curved shape $(b)$. The origin of the curvature is mainly due to the entropic penalty of the overhang sequences on a flat surface $(c)$, compared to a curved surface $(d)$ (shown schematically as a side view of an overhang). The overhangs on the curved surface have more accessible conformational space (shown in red) than they do on the flat surface.}    \label{fig:abs_fig}
    \end{figure}

We use the oxDNA model to study the effects of $2$D DNA origami deformation induced by the presence of single-stranded and double-stranded overhangs. An enhanced sampling method was utilized to sample the conformational space of the structure. We show that bent conformations are significantly enhanced as longer or denser overhangs are attached to the origami (Fig.~\ref{fig:abs_fig}). We show that the effect is of entropic origin, with the highly bent conformation being more favorable for structures with longer and denser overhangs. The results have implications both for DNA origami designs used for cargo delivery or surface-bound strand displacement based computation \cite{thubagere2017cargo,chao2019solving}, as the attached overhangs can have unintended effects on the structures' conformational ensemble. At the same time, the mechanism of entropy-induced curvature of DNA origami tiles can be exploited to impose certain preferred shapes to DNA structures for example as a design feature that is intended to distinguish between inside and outside in a bent structure, e.g. in the case of nanotubes \cite{li2022leakless}. We study here the effects of overhang length and density at different temperatures and salt concentrations, and show how the distribution of different shapes of the DNA tile is affected by their presence.


\section{Results and Discussion}
\subsection{Studied Systems}
      \begin{figure}
       \centering
     \includegraphics[width=0.5\textwidth]{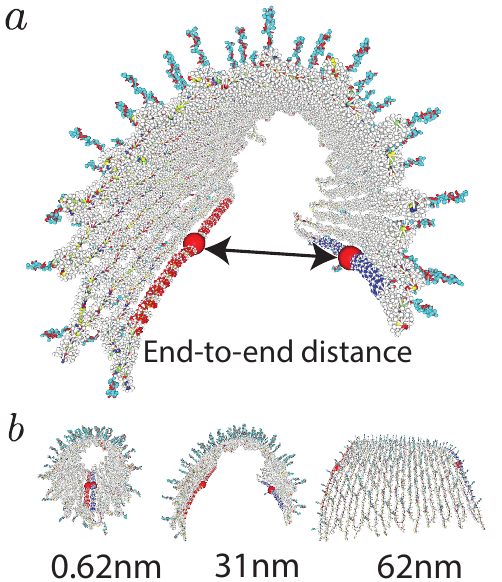}
    \caption{
    $(a)$ Mean structure of twist-corrected rectangular origami with $169$ overhang extensions comprised of twenty nucleotide bases. The arrows indicate the measured end-to-end distance order parameter ($R_{\rm ee}$) used to model the curvature of the structures. Low $R_{\rm ee}$ values correspond to high curvatures and high $R_{\rm ee}$'s to low curvatures. $(b)$ Mean structures of umbrella simulation windows equilibrated around $0.62$ nm, $31$ nm, and $62$ nm respectively.}
    \label{fig:op_rep}
    \end{figure}

\begin{figure*}[!ht]
     \centering
     \includegraphics[width=\textwidth]{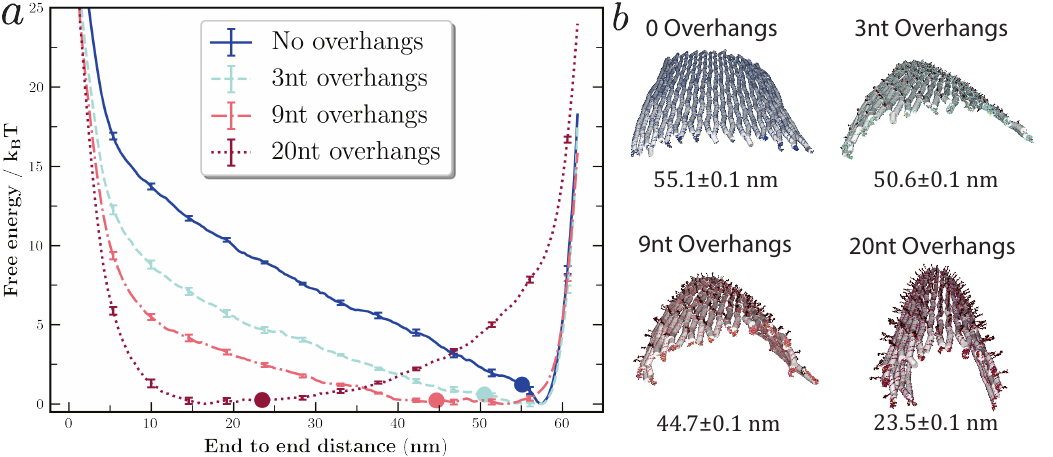}
    \caption{Effect of overhang length on structural curvature. $(a)$ Free-energy profiles as a function of the end-to-end distance of the twist-corrected rectangular tile origami. The dots indicate the location of the weighted average value. $(b)$ oxView visualization of mean structures and corresponding weighted average values. We compared structures with no overhangs,  $3$nt,  $9$nt, and $20$nt overhangs. The free energy profiles show that structures with a greater number of nucleotides in the overhangs, effectively longer overhang lengths, exhibit higher probabilities to be in states with higher magnitudes of curvature.}    \label{fig:length}
    \end{figure*}
    
The primary model system we chose to study the structural impact of DNA origami functionalization through extending staple strands was the twist-corrected rectangular DNA origami from Ref.~\cite{jungmann2016quantitative} (shown on the left in Fig.~\ref{fig:abs_fig}$a$). The DNA origami rectangles of this type have been utilized in various applications as a molecular canvas for nanometer scale positioning due to the $\sim5$ nm resolution of the site specific addressable DNA overhangs \cite{jungmann2016quantitative}.
We extended DNA overhangs from the $5'$ ends of the origami's staple strands, using a multitude of overhang conditions including single-strands, double-strands ($169$ overhangs, $85$ overhangs), as well as varying their length. To model experimentally realized systems, we initially created structures with nine nucleotide ($9$nt) and twenty nucleotide ($20$nt) long overhang extensions. The $9$nt overhang structure models the qPaint docking strand system from Ref.~\cite{jungmann2016quantitative} and the $20$nt structure represents the molecular positioning system created by Gopinath and collaborators in Ref.~\cite{gopinath2021absolute}.

Initial unbiased oxDNA simulations indicated the presence of structural curvature upon extension of overhangs, as can be seen in the calculated mean structure (Fig.~\ref{fig:op_rep}$a$). We next employed umbrella sampling to quantify the magnitude of curvature in rectangular origami structures with overhangs.  Umbrella sampling (US) is an enhanced sampling technique that allows us to efficiently sample all values along our chosen order parameter (OP) by introducing an external harmonic potential that biases the simulation to sample all desired states of the OP that represents different conformations of the structure (see Methods) \cite{kastner2011umbrella,snodin2019coarse,wong2022characterizing, wong2022free}. To model structural curvature, we chose our OP to be the distance between the centers of mass (COM) of the origami's long edges (see Fig.~\ref{fig:op_rep}), titled as the end-to-end distance ($R_{\rm ee}$), following the approach from Ref.~\cite{wong2022characterizing}.

The simulations allow us to assign probabilities to observe particular values of our OP in the conformational ensemble, which we use to quantify the rectangular origami's preference to exhibit different magnitudes of curvature as a function of the end-to-end distance between edges of the origami ($R_{\rm ee}$), as it represents a one-dimensional description of the curvature. 
The $R_{\rm ee}$ was sampled from $0.62$ nm to $62$ nm.

In addition to the $2$D origami from Ref.~\cite{jungmann2016quantitative}, we used oxDNA simulations (with similar OP choice, see Methods) to study the bending of the anti-parallel double layer rectangular origami from the work of Thubagere el al. \cite{thubagere2017cargo} and the six helix bundle rectangular origami from Dong et al. \cite{dong2021dna}. We studied their conformations both with and without 20nt long overhangs (see Fig.~\ref{fig:structures}). Furthermore, we created a modified $2$D origami rectangle able to exhibit overhang extensions from both sides. Lastly, we simulated another custom rectangular origami with 152 monomeric streptavidin proteins loaded onto the overhang extensions.

To compare the curvature exhibited by different structures, for all the studied designs we plot the free energy as a function of the end-to-end distance ($R_{\rm ee}$), obtained from the probability $p(R_{\rm ee})$ as $F(R_{\rm ee}) / k_{\rm B} T = -\ln{{p(R_{\rm ee})}} + C$, where we set constant $C$ such that $F(R_{\rm ee})$ is equal to 0 for the most probable value of $R_{\rm ee}$ (i.e. in its minima)\cite{doye2022free}. For each studied system, we also highlight in the plot (as a colored dot) the weighted average end-to-end distance $\langle R_{\rm ee} \rangle_p = \sum_{R_{\rm ee}^i} p(R_{\rm ee}^i) R_{\rm ee}^i$, where  $R_{\rm ee}^i$ are all the binned values of the end-to-end distance of $R_{\rm ee}$ that were sampled during the simulation.
    

\subsection{Effects of Overhang Length}

To investigate the impact of different lengths of overhang extensions on inducing the rectangular origami to curve, we simulated structures with varying numbers of nucleotides in their single-stranded poly-T overhangs. Free-energy profiles were computed for a rectangular structure with no overhangs, and then for three nucleotides, nine nucleotides, and twenty nucleotide-long overhangs respectively. All systems considered had the same density of overhangs, totaling $169$ overhangs attached to the $2$D origami tile.
We observed that the umbrella sampling simulation results show significant curvature in all origami structures with added overhang extensions. With the increasing length of the overhang, the probability of exhibiting a greater magnitude of curvature also increased (Fig.~\ref{fig:length}).


The rectangular structure with zero overhangs showed a weighted average $R_{\rm ee}$ value of $55.1 \pm 0.1$ nm, while the $3$nt, $9$nt, and $20$nt overhang structures had progressively smaller weighted average values (shown as colored dots in Fig.~\ref{fig:length}). These values quantitatively demonstrate a significant difference in the average curvature of the four different structures, where longer overhangs lead to increased curvature. In addition to computing free-energy profiles where the $R_{\rm ee}$ corresponded to states with the overhangs facing in an outward direction relative to the curvature, we also computed free-energy profiles of structures with the overhang extensions on the interior of the curvature (Supp.~Mat.~Fig.~S10). These profiles show that extending overhangs on one side makes it unlikely for the rectangle to conform to a state where the overhangs are on the interior.

The flexibility of a structure can be further seen from the differences between end-to-end distance free-energy profiles for the respective overhang lengths studied (Fig~\ref{fig:length}). From the relative flatness of the free-energy profiles, it can be seen that the structures have immense flexibility in exploring conformations outside of the minimum.
For example, the $2$D origami with $9$nt overhangs has a free energy minimum at $R_{\rm ee} = 48$ nm, but is only about $7$ times less likely to visit $R_{\rm ee}$ values of $30$ nm and $150$ times less likely to visit $R_{\rm ee}$ values of $7$ nm. Hence, it is expected to sample these conformations frequently in the bulk. Thus, to understand how flexible DNA origami will behave in solution and how likely certain conformations are, it is vital to obtain more accurate and complete information than static structural properties (e.g. AFM image) can provide us. 






\subsection{Entropy as the Driving Force}

To study the cause behind the induction of curvature through the addition of overhang extensions, we performed simulations with a modified oxDNA model where non-bonded interactions were turned off between specific groups of nucleotides for the origami with $9$nt and $20$nt long single-stranded overhangs. We expect that the excluded volume interactions (two strands cannot occupy the same space simultaneously) are the primary driving force behind the observed curvature. The single-stranded overhangs need to avoid overlapping with each other. The likelihood that two neighboring overhangs will clash is lower if the $2$D origami is curved.
Furthermore, the volume accessible to an overhang strand (Fig.~\ref{fig:abs_fig}$c$) will increase (Fig.~\ref{fig:abs_fig}$d$) if the $2$D origami bends more, making the bent configuration more favorable for entropic reasons, as the overhangs will have larger conformational volume accessible. The volume accessible to single-stranded DNA overhangs was also previously shown to be able to modulate the free-energy landscape of a DNA hinge nanostructure \cite{shi2020free}.

\begin{figure}
     \centering
     \includegraphics[width=0.48\textwidth]{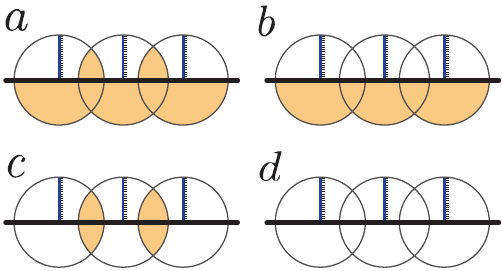}
    \caption{A schematic illustration of modified interaction potentials in the oxDNA models to test the effects of overhangs on DNA tile conformations: $(a)$ Non-modified interactions: the overhangs cannot occupy the same space at the same time due to excluded volume interaction (regions where the overhangs can "clash" with each other or the DNA tile highlighted in orange). $(b)$ A modified interaction where we switch off interactions between all pairs of DNA overhangs. $(c)$ A modification of the interactions where we switch off interactions between the overhangs and the nucleotides that are part of the 2D DNA origami tile. $(d)$ A modification where we switch off both interactions between pairs of overhangs, as well as between overhangs and nucleotides belonging to the the tile. The switching off of different interactions is not possible in actual experimental systems, but can be easily done in oxDNA model to identify the interactions' contribution to the observed bending effect.}    \label{fig:spheres}
    \end{figure}
 
 To study these phenomena, we simulated three different system modifications to decompose the underlying effects of the curvature (schematically illustrated in Fig.~\ref{fig:spheres}). In the first modification, we explicitly switched off all interactions between all pairs of overhangs, allowing the overhangs to pass through each other (Fig.~\ref{fig:spheres}$b$). For the second modification, we switched off interactions between the nucleotides in the single-stranded overhang extensions and the nucleotides in the rectangular origami's base, allowing the overhangs to freely pass through the origami surface (Fig.~\ref{fig:spheres}$c$). Finally, in the third modification, we combined both prior approaches and switched off the interactions between the overhangs with the rectangular tile as well as interactions with other overhangs (Fig.~\ref{fig:spheres}$d$). 

\begin{figure*}[!ht]
     \centering
     \includegraphics[width=\textwidth]{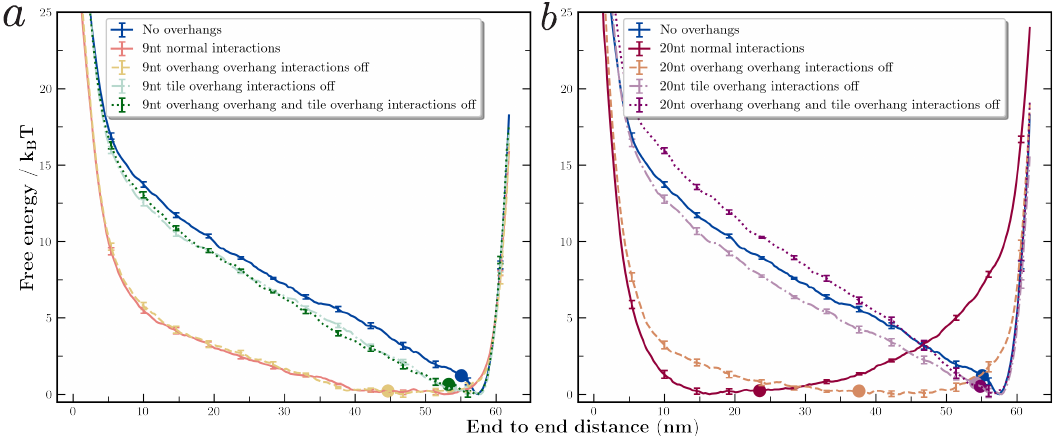}
    \caption{Quantification of the entropy effects on the curvature. Non-bonded interaction potentials of the overhang extensions for the $(a)$ $9$nt and $(b)$ $20$nt systems were modified to decompose the source of curvature. When the interactions of the overhangs with the rectangular tile are switched off the curvature of both systems closely mirrors the zero overhang curvature, indicating this interaction to be the main source of curvature. When the overhangs interactions with other overhangs are turned off the $20$nt system has a moderate change in curvature while the $9$nt system does not change, leading to the conclusion this condition is impactful only for long overhangs.}    \label{fig:entropy}
    \end{figure*}

For each of the modified simulations, we ran umbrella sampling simulations to reconstruct the free-energy profile of the end-to-end distance $R_{\rm ee}$ for the systems with $9$nt and $20$nt long single-stranded overhangs (Fig.~\ref{fig:entropy}).

\setlength{\belowcaptionskip}{-10pt}


Analyzing the modified simulations for the $20$nt overhangs structure, the free-energy profile for the first modification (interaction of overhangs with other overhangs switched off) moderately shifts the weighted average to lower curvature values. The second modification (overhangs interaction with DNA origami surface switched off) heavily increased the average $R_{\rm ee}$,  corresponding to greatly reduced curvature. The third condition (interaction with other overhangs and DNA origami surface switched off) has a weighted average nearly identical to the flat DNA origami with no overhangs (Fig.~\ref{fig:entropy}$a$). 

 The change in the $R_{\rm ee}$ value for the $9$nt long overhangs shows a different trend. For the first modification (no interactions between overhangs), the free-energy profile remains nearly identical to the case where no modification is introduced, implying that for these short overhangs, the effects of overhangs interacting through excluded volume are negligible. The second modification (switching off interactions between overhangs and DNA origami surface) shows a moderate increases in the average $R_{\rm ee}$, corresponding to a decrease in curvature. Thus, for shorter overhangs, the entropic origin of curvature appears to be due to the increased conformational space of individual overhangs, rather than due to the clashes with other overhangs.
 
Overall, these results show that the interactions between the overhang extension with the rectangular tile structure is the main contributing factor in causing the rectangular origami to curve. By adopting a curved surface, the number of conformational states that the overhangs extensions can explore increases, and consequently the curved structure is entropically favored (Fig.~\ref{fig:abs_fig}$b$).  
In addition to the curvature caused by the interaction between the overhangs and the rectangular tile, the interactions between the overhangs themselves also influence the curvature, but the result is only observed for longer extensions. All studies using modified interactions to study the effect of overhangs were performed at 1M salt concentration, so there were no long-range electrostatic interactions present. The Debye-H\"uckel length of the electrostatic potential in the model at 1M salt is about equal to the radius of the excluded volume of each backbone site, making the dominant interaction between the overhangs and the tiles just the excluded volume potential.

\subsection{Effects of Overhang Duplexes and Density}

To analyze the effect of overhang extensions in complex with their complementary strands and varying densities of overhangs, we designed origami structures with $169$ duplex overhangs ("dense" system),
"half-dense" structures with $85$ single-stranded overhangs,
and structures with $85$ duplex DNA overhangs ("half-dense" double-stranded). 
The double-stranded DNA overhang structures were designed to model a rectangular tile system in complex with complementary functionalized DNA strands. The half-density structures serve as a model for a system requiring further precision in the placement of overhand extensions, leading to a reduced number of rationally placed overhangs.

The free-energy profiles (see Supp.~Mat.~Fig.~S13) show that the formation of DNA duplex overhangs increases the curvature of both the $20$nt and $9$nt duplex overhang structures relative to their single stranded counterparts. While the free-energy minima of the $20$nt duplex overhang structure does not change compared to the $20$nt single-stranded overhang structure, an effective overall decrease in curvature can be seen from the shift in the weighted average value.



We further observed that when the density of the overhangs was approximately halved from $169$ to $85$, the curvature of the structure decreased (Supp.~Mat.~Fig.~s13), consistent with the fact that the entropic advantage of bending the origami will be lower for smaller number of the overhangs.   
We next compared the double-stranded overhangs and single-stranded overhangs for the half-density systems with $85$ overhangs. As observed for the dense system with $20$nt nucleotide overhangs, 
the $20$nt duplex overhangs structure had increased curvature compared to the single-stranded ones. However, we observed that the $9$nt duplex overhangs' mean end-to-end distance is slightly higher (indicating a less curved structure) than for the $9$nt single-stranded overhangs, which is the opposite effect from what we observed for the dense system with $169$ overhangs.   

The duplex overhangs behave more like wider stiff rods, whereas single-stranded DNA overhangs behave as freely-jointed chains with excluded volume. Thus, we expect the entropic penalty due to overhangs clashing with other overhangs to increase with double-stranded overhangs compared to single-stranded, an effect which is more pronounced for the longer ($20$nt) overhangs. However, the fact that the structure with $9$nt single-stranded overhangs is slightly more curved than DNA origami with $9$nt duplex overhangs, is indicative that the duplex overhangs are clashing less with the DNA origami surface than the single-stranded ones.

The sequence effects of the single-stranded regions was investigated by using the sequence-dependent version of the oxDNA model \cite{snodin2015introducing}, which has parameterized stronger Adenine-Adenine stacking compared to Thymine-Thymine stacking. The simulations (Supp.~Mat.~Fig.~S11) showed no significant difference in the curvature of the structures when the nucleotide sequence was Thymine or Adenine, indicating that the greater tendency for Adenines to stack compared to Thymines does not affect the origami's structural curvature.

Finally, to rule out the possibility that the induced curvature resulted from the arbitrary addition of overhang extensions to the structure, as opposed the addition of overhangs to a single side, we ran simulations of a $2$D rectangular origami with varied overhang extension placement. Namely, we ran simulations with overhangs on both sides, the top side, and the bottom side. As our previous used design disallows double sided overhangs as nearly all 5` and 3` ends terminate facing a single side, we modified the structure in oxView to enable double sided overhangs.
The simulation results show (Supp.~Mat.~Fig.~S15) that the addition of overhangs to both sides of the structure indeed results in a flat sheet. In addition, when overhangs were added to the bottom side of the rectangle it curved upwards, and when overhangs were added to the top the structure curved downwards.



\subsection{Effects of Temperature and Salt Concentration}

In all the cases considered above, we fixed the temperature at $20^{\circ}$C and the salt concentration in the oxDNA model at $1$M. To understand how the curvature of the rectangular tile origami might respond to different temperatures and salt concentrations, we computed free-energy profiles of end-to-end distance at four different temperatures and salt concentrations for the $20$nt long single-stranded overhang and no-overhang DNA origami structures. Simulations were run at $10^{\circ}$C, $20^{\circ}$C, $30^{\circ}$C, and $40^{\circ}$C, as well as at salt concentrations of $0.2$ M, $0.6$ M, $0.8$ M, and $1$ M (Supp.~Mat.~Fig.~S14)).

The free-energy profiles of end-to-end distance, and hence the curvature of the DNA origami, changes only slightly between different temperatures, corroborating the entropic origin of the curvature. 

When the salt concentration is varied, we observe that as the salt concentration decreases, the curvature of the structure also decreases. Our oxDNA model implements salt effects using the Debye-H\"uckel potential \cite{snodin2015introducing}, where the backbone sites of the nucleotides in the model interact with repulsive interaction with an effective charge. With decreasing salt concentration, the effective excluded volume occupied by the overhangs increases. While the curvature does not change drastically from $1$ to $0.6$ M salt, we observe moderately less curvature for $0.2$ M salt. The change in curvature for $0.2$ M salt as compared to other salt concentrations can be attributed to the exponential increase in the Debye length as the salt concentration approaches $0$. The increased electrostatic repulsion between the nucleotide backbones in the origami helices effectively flattens the structure, as can be seen for system with no overhangs (Supp.~Mat.~Fig.~S14). For the 20nt overhang lengths, while they have larger excluded volume at lower salts, the flattening of the origami at low salt concentrations still leads to larger average $R_{\rm ee}$ (smaller curvature) at lower salt concentrations. 


\subsection{Anti-Parallel Double Layer and Six Helix Bundle Rectangles}
\begin{figure*}
     \centering
     \includegraphics[width=\textwidth]{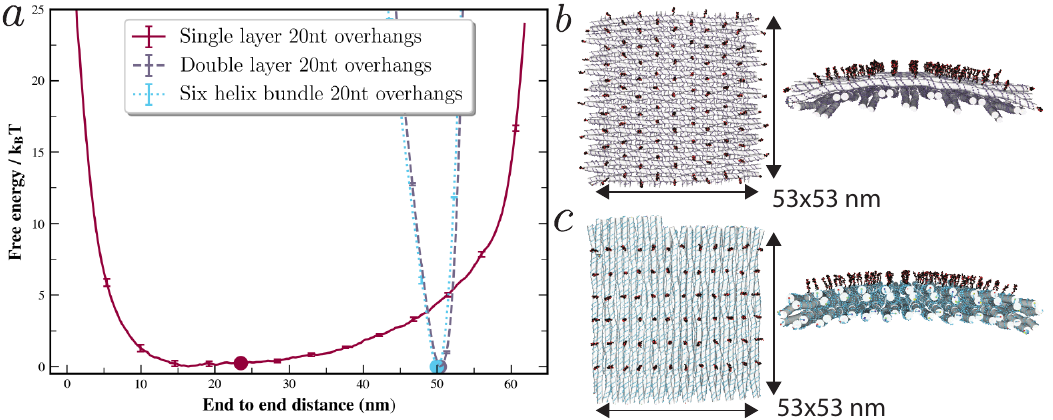}
    \caption{$(a)$ Free-energy profiles of $3$D rectangular origami structures. $(b)$ The double layer anti-parallel $20$nt overhangs and $(c)$ six helix bundle $20$nt overhangs structures show a remarkable increase in rigidity compared to the $2$D rectangular structure. The anti-parallel and six helix bundle free energy profiles show a much steeper rise in free energy upon perturbations from the free energy minimum indicating low flexibility, with weighted averages at $50.5$ nm and $50.0$ nm respectively.}    \label{fig:structures}
\end{figure*}

To compare the effects overhang extensions have on $3$D rectangular structures as opposed to the $2$D structure, we simulated two rectangular origami planes with 3D architectures. One structure consists of two anti-parallel $2$D rectangles (Fig.~\ref{fig:structures}$b$), while the other is a rectangle made of six helix bundles (Fig.~\ref{fig:structures}$c$) \cite{thubagere2017cargo, dong2021dna}. We extended twenty nucleotide overhangs from the staple strands of both systems.

The free-energy profiles of both $3$D rectangular structures show a significant increase in structural rigidity. The steepness of the free-energy profiles around the free-energy minimum, compared to the single layer $2$D rectangular origami, indicates that the end-to-end distance between edges of the structures are nearly constant, with huge penalty for bending. Additionally, the anti-parallel rectangle shows a stronger resistance to curvature compared to the six helix bundle structure (Supp.~Mat.~Fig.~S12). 
The anti-parallel structure (Fig.~\ref{fig:structures}$b$) has a free-energy minimum end-to-end distance at $50.5$ nm and weighted average of $50.5 $ nm. Similarly the six helix bundle (Fig.~\ref{fig:structures}$c$) has a minimum of $50.0$ nm and a weighted average of $50.0$ nm. In comparison, the single-layer DNA origami with $20$nt overhangs studied in previous sections has an end-to-end distance minimum at $16.4$ nm and a weighted average of $23.5 $nm, due to an underlying skewed distribution with a higher probability for states beside the minimum. In contrast, the measured values for $3$D double-layered structures indicate a rigid body with a large free-energy penalty for perturbations away from the minimum.  We note that these $3$D designs are a preferred choice for applications where flexibility of the structure should be minimal, preventing cross-talk between molecules attached to the structure's surface \cite{thubagere2017cargo}.

 \subsection{Experimental study of the overhang-induced bending}

\begin{figure*}[!ht]
     \centering
     \includegraphics[width=\textwidth]{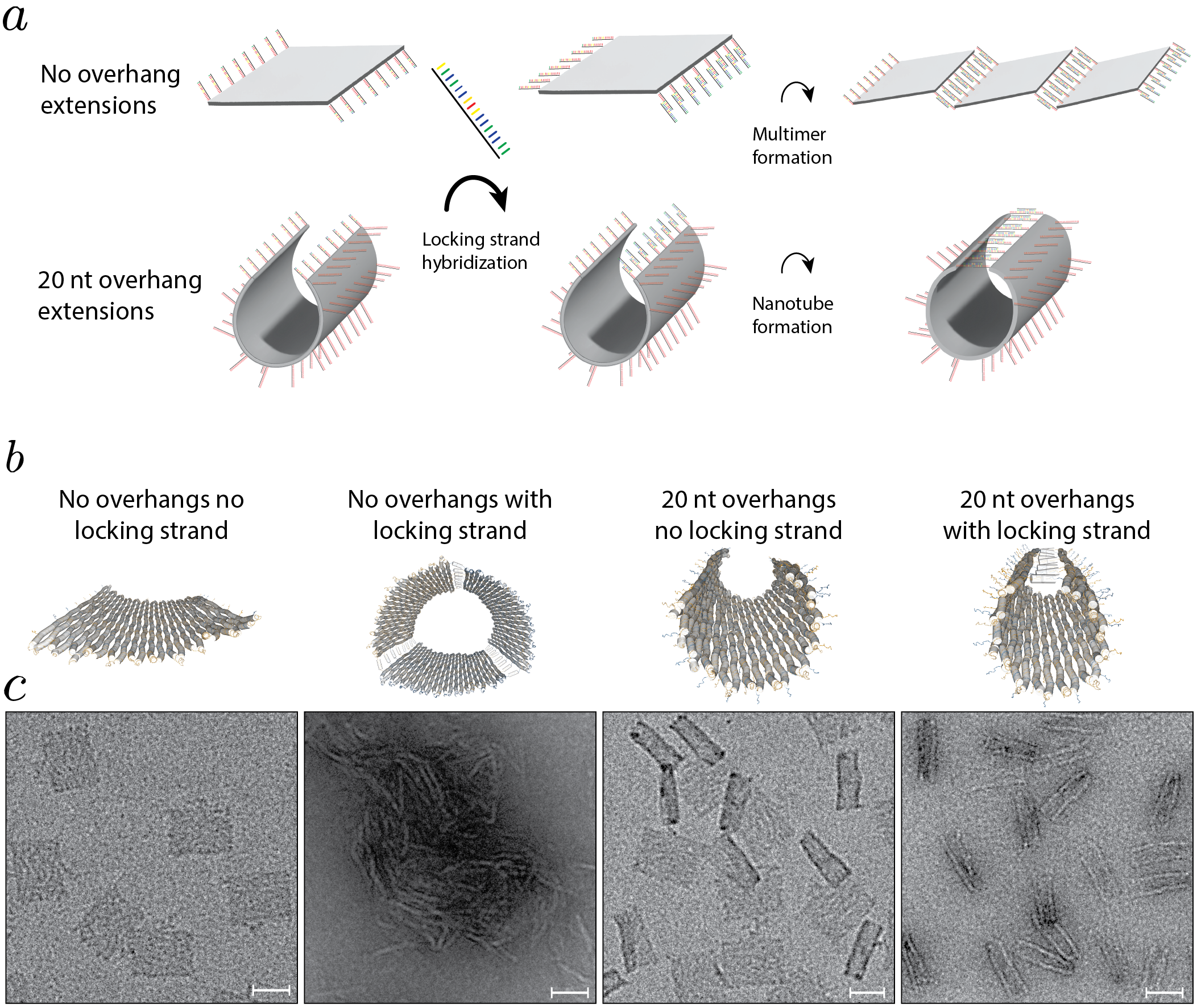}
    \caption{$(a)$ Schematic representation of experimental hypothesis. The top row illustrates the mechanism of multimers and aggregates formation, which are the most probable outcomes of locking strand hybridization for a twist corrected rectangular DNA origami with no overhangs, except those on the edges designed for locking strand hybridization. The bottom row depicts the formation of nanotubes, which are the most likely outcome of locking strand hybridization of a twist corrected rectangular DNA origami with densely placed 20nt overhang extensions. $(b)$  Four conditions are shown using oxDNA mean structures of rectangles with lockable edges:  no overhangs , no overhangs with lock strand resulting in a representative multimer (i.e. trimer), 20nt overhang extensions, and 20nt overhang extensions with a lock strand. $(c)$ TEM images of aforementioned four conditions (Scale bar: 50 nm)}    \label{fig:experimental_results}
    \end{figure*}

\begin{figure*}[!ht]
     \centering
     \includegraphics[width=\textwidth]{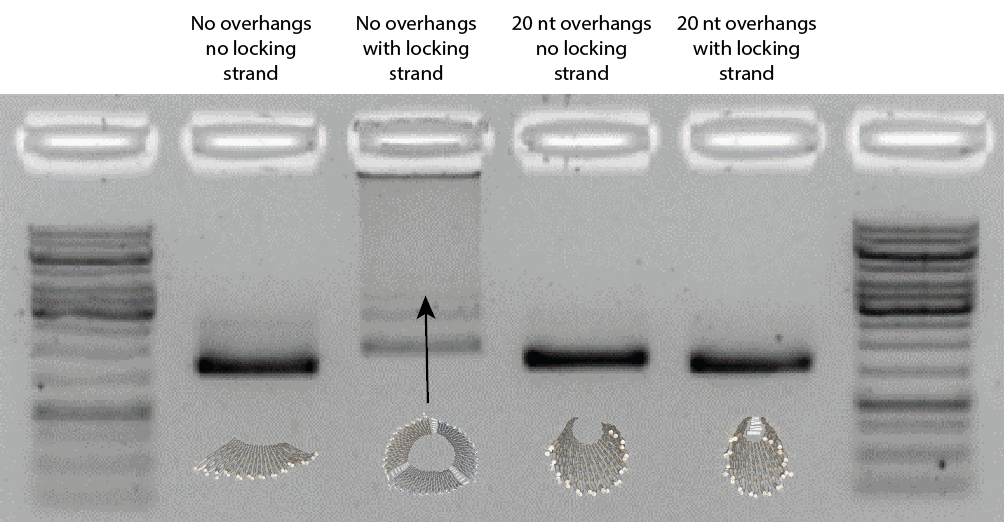}
    \caption{Image of gel electrophoresis experiment featuring four conditions of DNA origami rectangles with lockable edges:  no overhangs , no overhangs with a locking strand, 20nt overhang extensions, and 20nt overhang extensions with a locking strand. Corresponding oxDNA mean structures are overlaid below the well. For the second condition, one possible multimer (trimer) is shown, with an arrow pointing to the trimer band on the gel. The gel revels the formation of multimers and aggregate in the condition with no overhangs and a locking strand, while monomers are observed in all other conditions.}      \label{fig:gel}
    \end{figure*}

 To validate the phenomenon of overhang-induced curvature in flexible DNA origami structures, as revealed by our umbrella sampling simulations, we performed experiments for qualitative confirmation. We hypothesized that adding overhang extensions to the long ends of the rectangle--extensions that could hybridize to form a nanotube in the presence of a locking strand--the 20nt overhang rectangle would preferentially form nanotubes. In contrast, a rectangle without overhangs (except the lockable ones) would more likely form multimers and aggregates(Fig.~\ref{fig:experimental_results}$a$). A high density of overhangs would induce curvature, bringing the two ends of the rectangle into close proximity and favoring nanotube formation, whereas a flat sheet without overhangs would tend to interact and bind with other nearby rectangular DNA origami structures.
    
Using oxView, we designed overhangs to extend from the edge of the twist corrected rectangular origami, and a $16$ nucleotide long locking strand to hybridize them together. Each lockable overhang contained a $6$ nucleotide poly-T spacer followed by $8$ nucleotides complementary to half of the locking strand. Using this design, we explored four conditions, both in simulation and experiments. The first condition involved a rectangle with only the necessary overhang extensions for locking, but without the addition of the locking strand. The second condition using the same structure but included the locking strand. The final two conditions utilized a rectangle with 20nt overhang extensions, where in one case the locking strand was not added and another where it was added.
 
The gels (Fig.~\ref{fig:gel}) show that the system with no additional overhangs, and without the locking strand, yielded the designed rectangular sheets, as evidenced by a single band. Upon adding the locking strand to this structure, we observed the formation of aggregates and multimers, seen as the aggregate band at the top the well and a smear with multiple bands indicating a gradient of migration speeds. One of the possible multimers, a trimer, was simulated and is represented below the corresponding lane. The rectangle with added 20nt overhang extensions also showed a monomer band. Importantly, when the locking strand was added to this 20nt overhang structure we only observed a monomer band. This final lane indicates that the overhang extensions indeed induced the rectangle to curve, such that when the locking strand was added the formation of a nanotube was the most probable state. This is in stark comparison to the structure without the overhang extensions which instead formed a distribution of aggregates and multimers.

 Following the gel electrophoresis experiment, we preformed TEM imaging for all four conditions to corroborate the gels results (Fig.~\ref{fig:experimental_results}$b$). The TEM images show agreement with our hypothesis: the addition of the locking strand led to the formation of closed tubes in the condition with 20nt overhang extension, whereas the rectangle without additional overhangs displayed aggregates. 
 The condition with 20nt overhangs but without the locking strand exhibited a distribution of curved and flat structures. This is in agreement with the free energy profile of the 20nt overhang structure, which indicated that the rectangular origami would exist in a distribution of states in solution. However, the TEM images showed a higher proportion of flat structures than would be anticipated based on the free energy profile. This discrepancy can be explained by considering that TEM imaging requires adhering the structure to a charged surface, which would naturally skew the distribution toward a higher likelihood of flat structures. 
 
 Two primary mechanisms can contribute to this phenomenon. The first is the electrostatic attraction between the positively charged TEM imaging surface and the negatively charged DNA backbone. This interaction would be at it strongest when the maximum surface area of the DNA origami is in contact with the surface. The most straightforward way to achieve this is by forcing the rectangular origami to adopt a flat conformation. Another way to increase surface area would be to have the overhang extensions lie flat on the surface. However, this would induce an entropic penalty associated with confining overhang extensions to a surface.  These two factors likely shift the distribution of state in favor of flat structures.
 

Next, we explored the effects of the salt type on the DNA origami tile overhang-induced bending. All our modeling was done with oxDNA model with monovalent salt conditions treated with Debye-H\"{u}ckel approximation, but our experiments described above have been done in a $12.5$ mM Mg$^{2+}$ buffer, typical conditions often used for a majority of DNA origami experiments. However, our simulations results should be valid for any experimental conditions with high salt concentration, where DNA backbone charges are highly screened. To verify this assumption, we have also repeated all experiments in high monovalent salt concentration (1M Na$^+$). We observed the same behavior as in the divalent salt buffer, and all experimental results are summarized in Supp.~Mat.~Section S1 and Supp.~Mat.~Fig.~S2 \& ~S5.

\subsection{Protein cargo-induced bending}
\begin{figure*}[!ht]
     \centering
     \includegraphics[width=\textwidth]{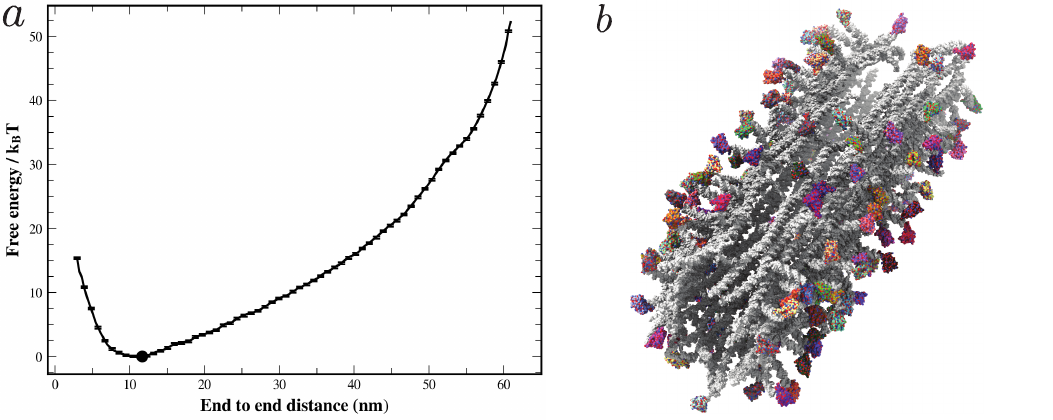}
    \caption{($a$) Free-energy profile of the origami loaded with 152 monomeric streptavidin proteins. ($b$) Electrostatic surface visualization of protein loaded origami. The proteins are colored randomly by strand. Loading the proteins onto the DNA nanostructure leads the structure to exhibit large curvature, with an $R_{\rm ee}$ of 11.6 nm.}    \label{fig:protein-conjugated}
    \end{figure*}
Finally, we studied whether the similar effect that is observed with DNA overhangs is also observed when single-stranded overhangs on DNA origami tile are attached to proteins. Using protein-patterned DNA origami is of increased interest in bionanotechnology, e.g. in diagnostic and therapeutic applications. For example, a certain protein pattern can be used to specifically enhance binding to a target cell which has a specific receptor pattern \cite{bila2022multivalent}. Our theoretical argument about entropy induced binding, as outlined for DNA overhangs in Fig~\ref{fig:abs_fig}, should also equally apply to proteins attached on the origami surface. To verify this assumption, we have performed coarse-grained simulation using ANM-oxDNA model \cite{procyk2021coarse}, an oxDNA model which also include proteins represented as beads (1 bead per aminoacid) connected by springs, which interact with excluded volume, thus providing an approximate rigid-body representation of proteins. As expected, we have observed in our simulation the same induced bending as in the case of single or double-stranded DNA overhangs (Fig.~\ref{fig:protein-conjugated}). The weighted average $R_{\rm ee}$ of the protein loaded structure with 20nt overhangs was 11.6 nm, more curved than 20nt duplex overhangs structure ($R_{\rm ee}=18.0$ nm). However, the protein surface charges are not captured in the ANM-oxDNA model, and to further verify that the assumption of charges screened by high salt concentration is valid, we have further performed experimental measurements, following the same experimental setup as used previously for the DNA overhangs. We verified the successful protein attachments by AFM imaging (Supp. Fig. S7 and S8). The gels then showed a consistent formation of just a monomer band for the locked DNA origami structure with the proteins (Supp. Fig. S3 and S4), e.g. the same behavior we observed in verification of induced curvature by DNA overhangs alone in Fig.~\ref{fig:gel}. TEM imaging of protein-functionalized DNA origami shows the formation of closed tubes only (Supp. Figs. S5 and S6). We have thus successfully verified that protein patterning of DNA origami surface can induce the same bending.



\section{Conclusions}
In this study we characterized both copmutationally and experimentally the effect of site addressable functionalizable overhang extensions on the structure of twist-corrected rectangular DNA origami. We showed that upon extension of overhangs from the end of the staple strands, there was a significant induced curvature in the structure. Furthermore, we quantified how the average curvature is influenced by different variables including the overhang length, single-stranded versus double-stranded overhangs, density of overhangs, overhang sequence, temperature and salt concentration. The insights provided by the computed free-energy profiles paint a robust picture of the dynamics of flexible DNA origami in solution. These results conclude that in order to rationally design flexible DNA nanostructures with nanometer scale precision, it is vital to take into account the effect that the extensions will have on modifying the resultant structure. 

While in the past experimental evidence indicate the existence of this phenomenon \cite{zhang2021prescribing, mohammed2013directing, cecconello2016dna, yu2023cytodirect}, this study was the first to rigorously quantify the degree of curvature induced by the functionalization of origami by single-stranded and double-stranded overhangs, with implications for designs with other guest molecules (such as proteins or gold nanoparticles), which we expect to show similar effects. We showed that as the number of nucleotides in the overhang extensions increases, the average curvature of the rectangular origami increases as well. In addition, we displayed how the curvature of the origami structure increased as a response to the formation of double stranded DNA overhangs.


By sampling the dynamics of an implicitly solvated DNA origami, our coarse-grained model, oxDNA, accurately predicts how entropic effects impact structural properties. Utilizing modified interaction potentials, we decomposed the entropic penalty caused by excluded body steric interactions. The increased state space accessible to the overhang extensions upon origami curvature results in a free energy landscape with a high probability of curved states, thus explaining the entropic origin of the observed DNA origami bending.

 
 Furthermore, when the simulations of DNA origami are utilized to calculate free-energy profiles though umbrella sampling simulations, we are able to obtain a much more robust characterization of not only the mean conformation of our nanostructure, but also information on all possible states, along with the associated probability, for a flexible DNA origami. 
Finally, we ran oxDNA molecular dynamics simulations of 3D anti-parallel double layer rectangular DNA Origami and a six helix bundle rectangular DNA Origami with 20nt overhang extensions. The simulations showed that there was a significant increase in structural rigidity with negligible deformation of the planar shape of both 3D nanostructures as compared to the 2D rectangular origami. Hence, 3D DNA origami rectangles would be recommended in an application where a planar structure is required or precise spatial resolution needs to be retained upon overhang functionalization. 

We verified our model prediction experimentally by designing a system that either forms tubes or multimeric sheets, depending on the presence or absence of induced bending by overhangs. Our experiments confirmed the simulation predictions, demonstrating experimentally the induced bending phenomena.


We have further explored the salt effects. Our simulations were performed using 0.2 to 1M salt, with electrostatic effects approximated in the model by Debye-H\"{u}ckel potential and excluded volume interactions. DNA origami experiments are typically done at high salt concentration (12 mM Mg$^{2+}$, but have also been done at magnesium-free high sodium buffer \cite{martin2012magnesium}), so the highly screened short-range electrostatic repulsion between DNA strands should be accurately captured by the model. Our experiments, performed both in magnesium and sodium buffers behaved as predicted, thus corroborating our simulation results.

Finally, we have also carried out a simulation study of DNA origami functionalized with attached proteins, to imitate the use of a DNA origami as a "pegboard" to position cargo in a specific pattern. Similarly to the bending induced by the overhangs, our simulations predict the same entropy-induced bending phenomena to occur due to the excluded volume occupied by the attached proteins. We have also verified this model prediction experimentally.

Overall, our findings provide an important quantification of the effects of functionalization of the DNA origami on their conformational ensemble, with implications for functional designs for fields such as photonics, diagnostics and drug delivery. We note that the effect of origami curvature induced by overhangs can be potentially exploited to induce desired curved shape on a DNA origami surfaces. The typical conformation for all the structures that have been simulated are provided in the Supp.~Mat.~Fig.~S16 and summarized in Supp.~Mat.~Table S1. We provide the simulation code developed to effectively sample all the conformations as an open source software at \url{https://github.com/mlsample/hairygami_umbrella_sampling}.



\section{Materials and Methods}
\subsection{Simulations}
To simulate the structural dynamics of the DNA nanostructures, we use the oxDNA coarse-grained model. Specifically, we used oxDNA$2$ \cite{snodin2015introducing} implemented on GPU \cite{rovigatti2015comparison, poppleton2023oxdna}. The time-step used for all simulations was 15 fs, as previously used for origami characterization \cite{wong2022characterizing}. Simulations were performed at 293.15 K and 1 M salt concentration with averaged stacking and hydrogen bonding strength unless otherwise specified. An Anderson-like thermostat was used for temperature coupling and the salt concentration is modeled using the Debye-H\"uckel potential. A diffusion coefficient of 2.5 in simulation units was used, enabling us to sample longer timescales. The starting configurations were prepared by the oxView tool \cite{poppleton2020design,bohlin2022design}, where we extended a specified number of staple strands to the $5'$ end of the $2$D rectangular origami tile.
Umbrella sampling \cite{torrie1974monte, torrie1977nonphysical} was then used to compute all of the free energy profiles in this study. Adopting the approach from Ref.~\cite{wong2022characterizing}, we introduced a biasing harmonic potential $V_{\rm bias} = \frac{k}{2} (R_{\rm ee} - R_0^{w_i} )^2$, where $R_{\rm ee}$ is the distance between the centers of mass of the edges of the DNA origami, and $R_0^{w_i}$ is a variable that is selected for a particular simulation window. For each studied system, we simulated multiple independent windows, where a window is a standalone simulation that was run for $20$ million simulation steps. For each window $w_i$, we set a different value of $R_0^{w_i}$, ranging from $0.62$ to $62$ nm using $100$ simulation windows with the increment between each window being $0.62$ nm. The maximum order parameter value was $62$ nm, was chosen because the average $R_{\rm ee}$ of the rectangular structure when forced to be flat using a repulsion plane external potential was calculated to be about $60$ nm, and we aimed to profile the structure slightly beyond flat leading to our choice of $62$ nm as the maximum value. The minimum value of $0.62$ was picked as a result of our decision to use $100$ simulation windows, where $100$ windows starting at $62$ nm decreased by $0.62$ nm over our windows leads to a minimum value of $0.62$ nm. We chose $100$ simulation windows and the biasing harmonic potential with a spring constant $k = 11.418$ pN/nm to guarantee sufficient overlap between neighboring windows. The $R_{\rm ee}$ was fit to the angle between the center of the rectangle and the ends using polynomial fitting (Supp.~Mat.~Fig.~S9). Furthermore, we computed free energy profiles of structures with the overhang extensions on the interior of the curvature (Supp.~Mat.~Fig.~S10). 
The free-energy profiles which quantified the end to end distance of the rectangles where the overhangs were on the interior of the curvature (Supp.~Mat.~Fig.~S10) were setup manually in oxView by manipulating the nucleotides, and then relaxed using Monte Carlo sampling while applying external potentials to enforce all native hydrogen bonds and maintain the forced inverse curvature. The spring constant of the umbrella simulations were $k = 114.18$ pN/nm to enforce the highly unfavorable curvature.

Once the $100$ production windows were run, we used the Weighted histogram analysis method \cite{ferrenberg1989optimized, kumar1992weighted, wham} (WHAM) to unbias our simulations results and provide free energy values as a function of our order parameter. We chose to use $200$ bins for our WHAM analysis. The plotted free energy over $k_{\rm B} T$ is calculated by dividing the WHAM calculated free energy by the temperature. We used Ref.~\cite{wham} for Monte Carlo bootstrapping error analysis to create the error bars on the free energy values, which are also divided by the temperature of the system. Using the free-energy values calculated from WHAM, the weighted average of the end-to-end distance is then computed, using the probability of respective end-to-end values. The error of the weighted average was computed using parametric bootstrapping, modeling the free energy profiles as a multivariate Gaussian with means equal to the free energy values and standard deviations equal to the error provided by the WHAM Monte Carlo bootstrapping error analysis.  

The ANM-oxDNA model \cite{procyk2021coarse} was used to run the simulations of the rectangular origami with 152 monomeric streptavidin proteins loaded onto overhang extensions. The biotin linker was modeled using a skewed Gaussian potential, parameterized as done in Refs.~\cite{narayanan2022coarse, xu2023high}. All other parameters were the same previously mentioned, expect a time step of 10 fs used. The oxDNA analysis tools and Tacoxdna software packages were used to convert the simulated ANM-oxDNA structure into PDB format\cite{suma2019tacoxdna, poppleton2020design}. Then we used ChimeraX software to visualize the protein loaded origami PDB \cite{meng2023ucsf, pettersen2021ucsf, goddard2018ucsf}.

\subsection{Code Availability}
We make the computational methods and all the software tools developed here to setup umbrella sampling of conformations of nanostructures with oxDNA model freely available at
\url{https://github.com/mlsample/hairygami_umbrella_sampling}, along with examples to reproduce our results. We also provide an automated setup that enables running multiple instances of oxDNA molecular dynamics simulations on a single CUDA-enabled GPU card. With our setup, we were able to run in parallel $40$ oxDNA simulations of a single DNA origami on one NVIDIA A100 card. 
Running simulations in parallel on a single card is multiple times faster than running one simulation after other. We observed an increased throughput of $2$ times for $5$ simulations or $2.6$ times for $40$ simulations in parallel on a single GPU card. 
Hence, our setup opens a way for massively parallelized nanostructure characterization even with limited computational resources.
\subsection{Experimental setup}
The gel electrophoresis and negative staining TEM experiments to characterize the nanostructure assemblies are detailed in Supplementary Material Section S1. The staples for the nanostructure designs are listed in Table S2. Additionally, all the experimentally realized designs were uploaded to nanobase.org public repository \cite{poppleton2022nanobase} in the oxDNA file format.

\section{Acknowledgments}
We acknowledge support from the ONR Grant N000142012094 and DURIP ONR grant no.~N000142112876. This material is based upon work supported by the National Science Foundation under Grant No.~2227650. We further acknowledge use of the Extreme Science and Engineering Discovery Environment (XSEDE), which is supported by National Science Foundation grant number TG-BIO210009, as well as Research Computing at Arizona State University for providing HPC resources. We thank Jonathan Doye, Thomas Ouldridge, Paul Rothemund, and Matteo Guareschi for helpful discussions and to Joel Joseph for help with the design of the simulated structures. We thank Nicholas Stephanopoulos and Hao Yan for providing access to their experimental facilities. We thank the anonymous reviewer for their suggestion to improve the manuscript.

\bibliographystyle{old-mujstyl}
\bibliography{newrefs}

\end{document}


\renewcommand{\thepage}{S\arabic{page}} 
\renewcommand{\thesection}{S\arabic{section}}  
\renewcommand{\thetable}{S\arabic{table}}  
\renewcommand{\thefigure}{S\arabic{figure}}
\title{Supplementary Information:  Hairygami: Analysis of DNA Nanostructures' Conformational Change Driven by Functionalizable Overhangs}
\author{Matthew Sample}
\affiliation{School for Engineering of Matter, Transport, and Energy, Arizona State University, Tempe, AZ 85287, USA}
\affiliation{School of Molecular Sciences and Center for Molecular Design and Biomimetics, The Biodesign Institute, Arizona State University, 1001 South McAllister Avenue, Tempe, Arizona 85281, USA}
\author{Hao Liu}
\affiliation{School of Molecular Sciences and Center for Molecular Design and Biomimetics, The Biodesign Institute, Arizona State University, 1001 South McAllister Avenue, Tempe, Arizona 85281, USA}
\author{Thong Diep}
\affiliation{School of Molecular Sciences and Center for Molecular Design and Biomimetics, The Biodesign Institute, Arizona State University, 1001 South McAllister Avenue, Tempe, Arizona 85281, USA}
\author{Michael Matthies}
\affiliation{School of Molecular Sciences and Center for Molecular Design and Biomimetics, The Biodesign Institute, Arizona State University, 1001 South McAllister Avenue, Tempe, Arizona 85281, USA}
\affiliation{TU Munich, School of Natural Sciences, Department of Bioscience, Garching, Germany}
\author{Petr \v{S}ulc}
\affiliation{School of Molecular Sciences and Center for Molecular Design and Biomimetics, The Biodesign Institute, Arizona State University, 1001 South McAllister Avenue, Tempe, Arizona 85281, USA}
\affiliation{TU Munich, School of Natural Sciences, Department of Bioscience, Garching, Germany}

\maketitle

\setlength{\abovecaptionskip}{0pt}
\setlength{\belowcaptionskip}{0pt}

\section{Experimental methods}

\subsection{Origami Design and Assembly}

Detailed designs and sequences are shown in Figure S1 and Table S1. Briefly, 
rectangular DNA origami is designed by routing the M13mp18 Scaffold (Bayou Biolabs), with partial staples ready to be extended with poly-T overhangs to curve the origami plate. Edge handles for the responsive locking mechanism are designed on the two sides of the rectangle. The origami is assembled by mixing the 10 nM scaffold strand with the respective staple strand sets (with or without poly-T extension) at a 1:10 molar ratio, and specifically the designated handle strands at a 1:20 molar ratio, in 1x TAE-$\mathrm{Mg^{2+}}$ buffer (40 mM tris base, 20
mM acetic acid, 1 mM EDTA, and 12.5 mM magnesium acetate). Thermal annealing is executed as follows: 75 \degree C to 60 \degree C at a rate of 1 \degree C per 5 min, 60 \degree C to 40 \degree C at a rate of 0.1 \degree C per 3 min,  40 \degree C to 20 \degree C at a rate of 0.1 \degree C per 1 min, hold at 20 \degree C. 

For experiments involving sodium as the only source of cations, 1x TAE-$\mathrm{Na^{+}}$ buffer (40 mM tris base, 20mM acetic acid, 1 mM EDTA, and 1M sodium chloride) is used instead while all other conditions for origami assembly (e.g. concentrations, molar excess, annealing protocols) remain to be the same. 

Excess staple strands of the origami are purified through amicon ultrafiltration. Briefly, 100 kD amicon columns (MilliporeSigma) are passivated with fresh buffer followed by loading the diluted origami sample (to \SI{400}{\micro\liter}) to fill up the column. The centrifugation is done with a rate of 2k rcf followed by refilling the column with more buffer to wash the remaining staples. Such a procedure is done for overall 5 times to remove the staples thoroughly. Purified origami sample is collected by flipping the column upside down in an unused tube and doing a light centrifugation (3k rpm, 3 min). The concentration of the origami is determined by Nanodrop based on the absorbance at 260 nm. Unused samples are stored at 4 \degree C refrigerator until further processing. 

The locking strand is added in 50 times excess to the purified origami sample and the mixed sample is annealed to ensure the hybridization of the handles on the edges of the origami plate. The annealing protocol is set as follows: 37 \degree C incubation for 1 hour, 37 \degree C to 20 \degree C at a rate of 1 \degree C per 5 min.

\subsection{Protein conjugation}


Biotinylated strands are used to attach monomeric streptavidin (SAE0094, MilliporeSigma) to the poly-T extensions on rectangular DNA origami.

1.2 fold excess amount, against the number of the poly-T extensions, of biotinylated strands are added to the purified DNA origami sample (note 156 poly-T extensions per origami). The sample is then subjected to thermal annealing with the protocol included as follows: 37 \degree C incubation for 1 hour, 35 \degree C to 20 \degree C at a rate of 1 \degree C per 5 min. Furthermore, monomeric streptavidin with a 1.2 fold excess to biotinylated strands is further added, with the mixed sample incubated at 35 \degree C for 1 hour afterward. Lastly, locking strands with a 50-fold excess to the purified origami molecules are added to the protein-conjugated origami sample. The mixture undergoes thermal annealing with the same protocol listed above to lock the curved conformation for structural imaging. 

\subsection{Gel Electrophoresis}

Agarose gel (1.2\%, 1\% Invitrogen SYBR safe in 1x TAE-$\mathrm{Mg^{2+}}$ buffer) is used for the experiment of gel electrophoresis. Briefly, \SI{5}{\micro\liter} of the DNA origami sample annealed in 10 nM are loaded into the wells after mixing with the native loading buffer (50\% glycerol in 1x TAE-$\mathrm{Mg^{2+}}$ buffer). After running for 1.5 hour on ice bath at 90 V, the gel was imaged using Gel Doc XR+ system gel imager (Bio-Rad).

\subsection{Negative staining TEM}

\SI{5}{\micro\liter} of the DNA origami sample is absorbed onto the plasma charged (30s, Emitech k100x) Formvar carbon-coated copper grids. Excess liquid is blotted by the filter paper after 3 min, following with the staining using a 2\% aqueous uranyl acetate solutions for 30 s. For samples with the locking strand added, one additional buffer wash is executed between the sample incubation and staining steps to remove the excess DNA strands. After above procedures, the grids are allowed to dry under ambient conditions for at least 30 min before imaging. The TEM images are collected with a Thermo Scientific Talos L120C G2 TEM operated at an acceleration voltage of 120 kV.
\begin{figure*}[!ht]
     \centering
     \includegraphics[width=0.9\textwidth]{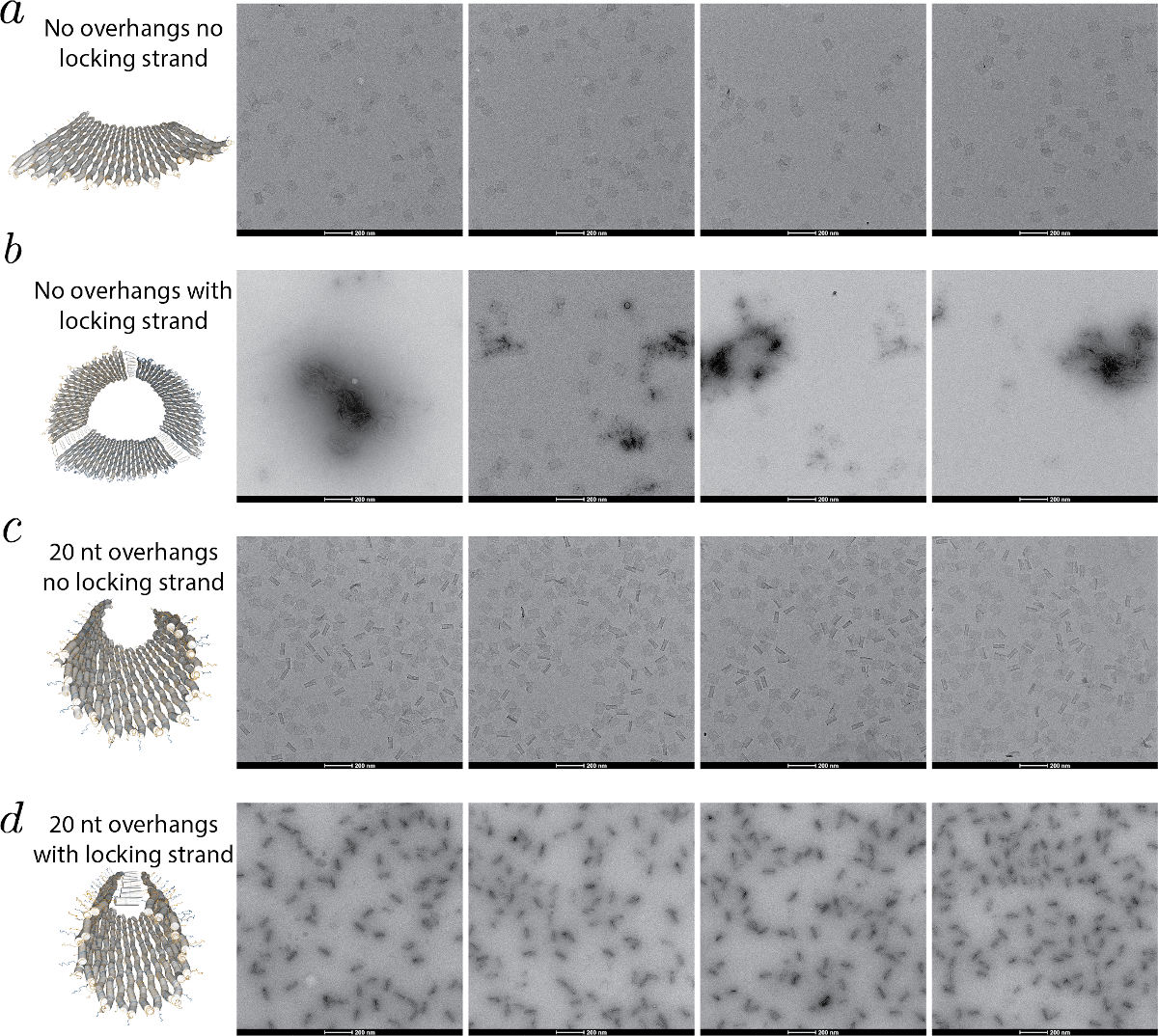}
    \caption{Negative staining TEM images of twist corrected DNA origami rectangles with $(a)$ no overhang and no locking strand, $(b)$ no overhangs with locking strands, leading to aggregates and multimers (representative trimer shown), $(c)$ 20nt overhang extensions with no locking strand, $(d)$ 20nt overhang extensions with locking strand.}    \label{fig:tem}
    \end{figure*}

\subsection{AFM imaging}

\SI{5}{\micro\liter} of the purified DNA origami sample (3 nM) is deposited onto a freshly cleaved mica surface (Ted Pella, Inc.) and incubated for 2 min. \SI{60}{\micro\liter} of 1x TAE-$\mathrm{Mg^{2+}}$ buffer is added onto the mica surface with \SI{5}{\micro\liter} of $\mathrm{NiCl_{2}}$ (200 mM) added subsequently to assist adsorption. The sample is then imaged in “ScanAsyst in Fluid” mode with a ScanAsyst-$\mathrm{liquid^{+}}$+ tip on the MultiMode 8 AFM (Bruker).

\begin{figure*}[!ht]
     \centering
     \includegraphics[width=0.9\textwidth]{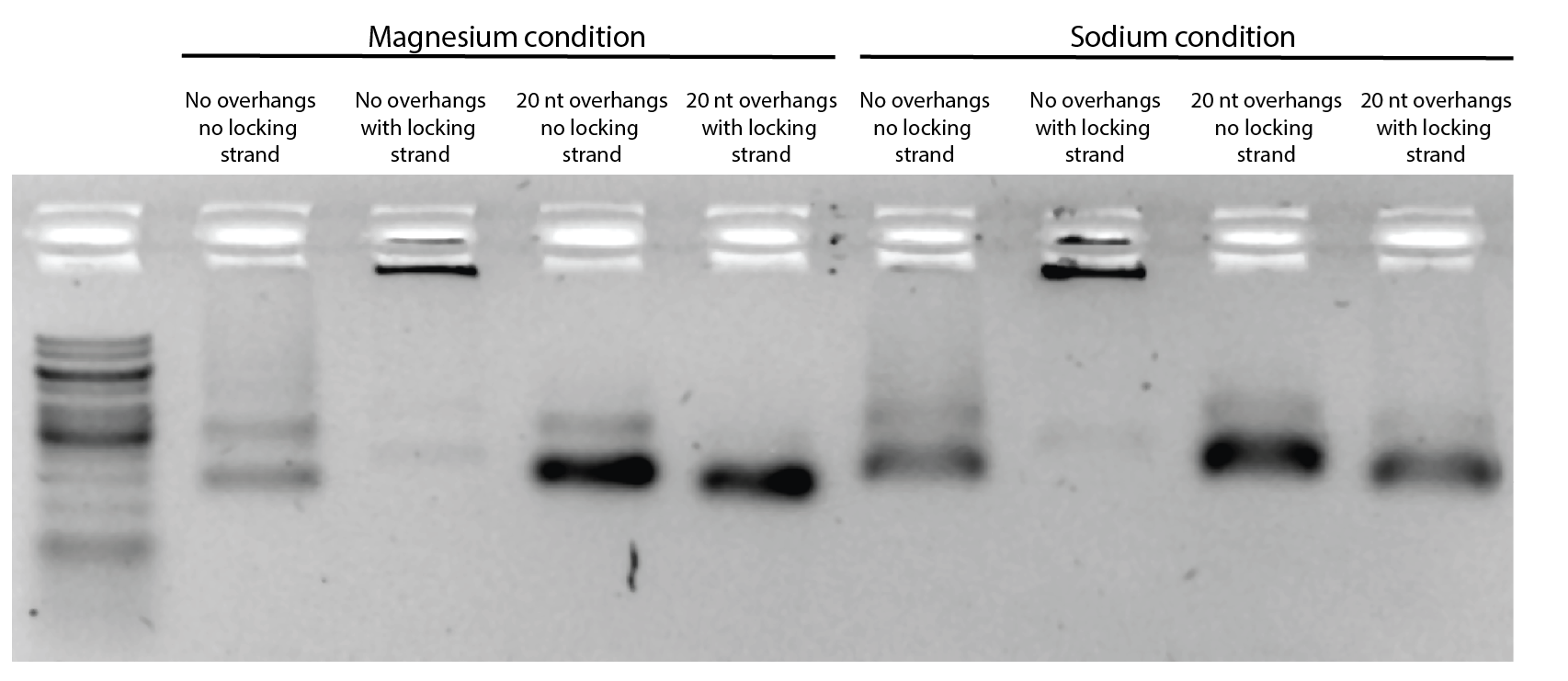}
    \caption{Agarose gel characterization of the cation effect in the buffer on DNA origami conformation change. The samples from the two conditions are subjected to the same thermal annealing for comparison. Similar results are observed for both conditions.  }    \label{fig:GelMgNa}
    \end{figure*}

\begin{figure*}[!ht]
     \centering
     \includegraphics[width=0.4\textwidth]{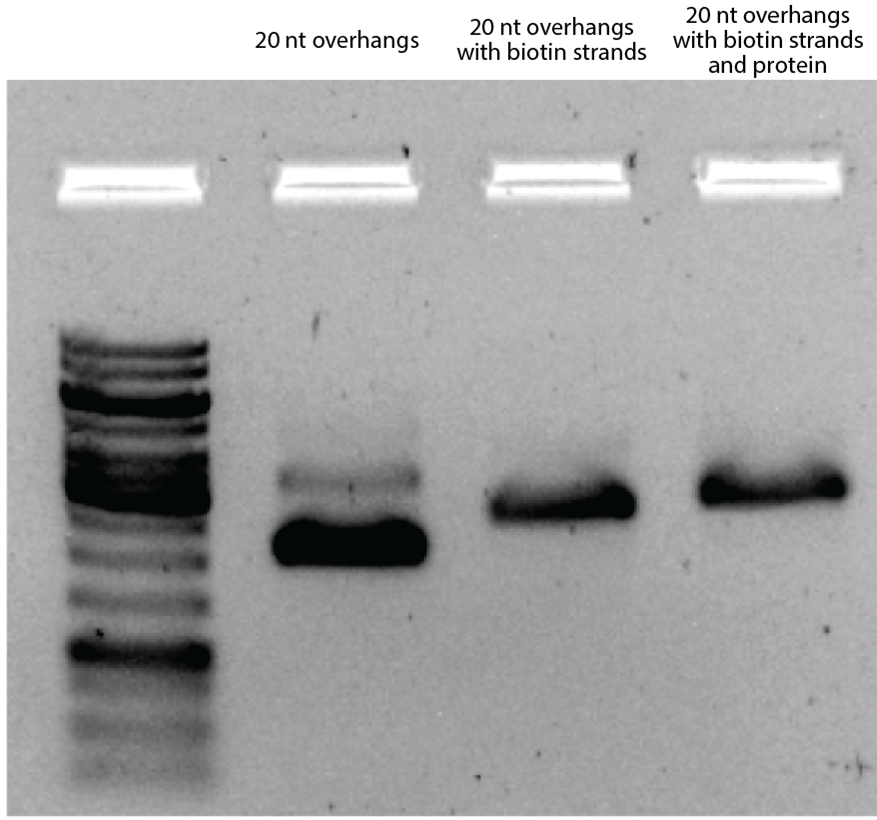}
    \caption{Gel image of the DNA origami rectangles with different modifications. Samples with just poly-T extensions, hybridized biotinylated linkers, and further monomeric streptavidin attachment exhibit different migration rates. }    
    \label{fig:GelproteinConj}
    \end{figure*}

\begin{figure*}[!ht]
     \centering
     \includegraphics[width=0.5\textwidth]{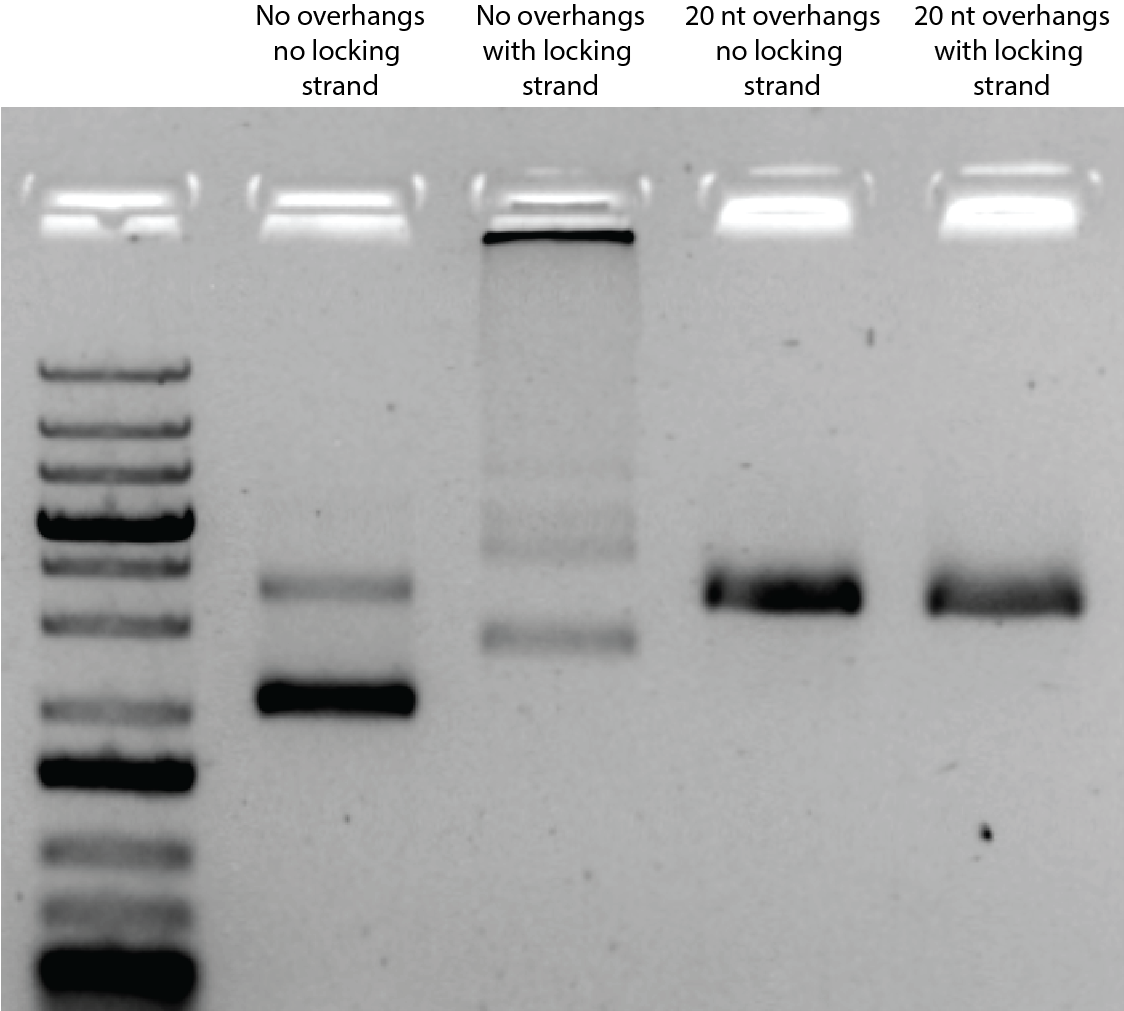}
    \caption{Agarose gel characterization of the DNA origami conformational change with attached proteins. Biotinylated strands and the protein are added under all conditions, and they are expected to attach specifically to the samples with extended poly-T overhangs only. A similar trend is observed compared to the experiment without the addition of biotinylated strands and the associated monomeric streptavidin.}    
    \label{fig:GelproteinCurve}
    \end{figure*}

\begin{figure*}[!ht]
     \centering
     \includegraphics[width=0.9\textwidth]{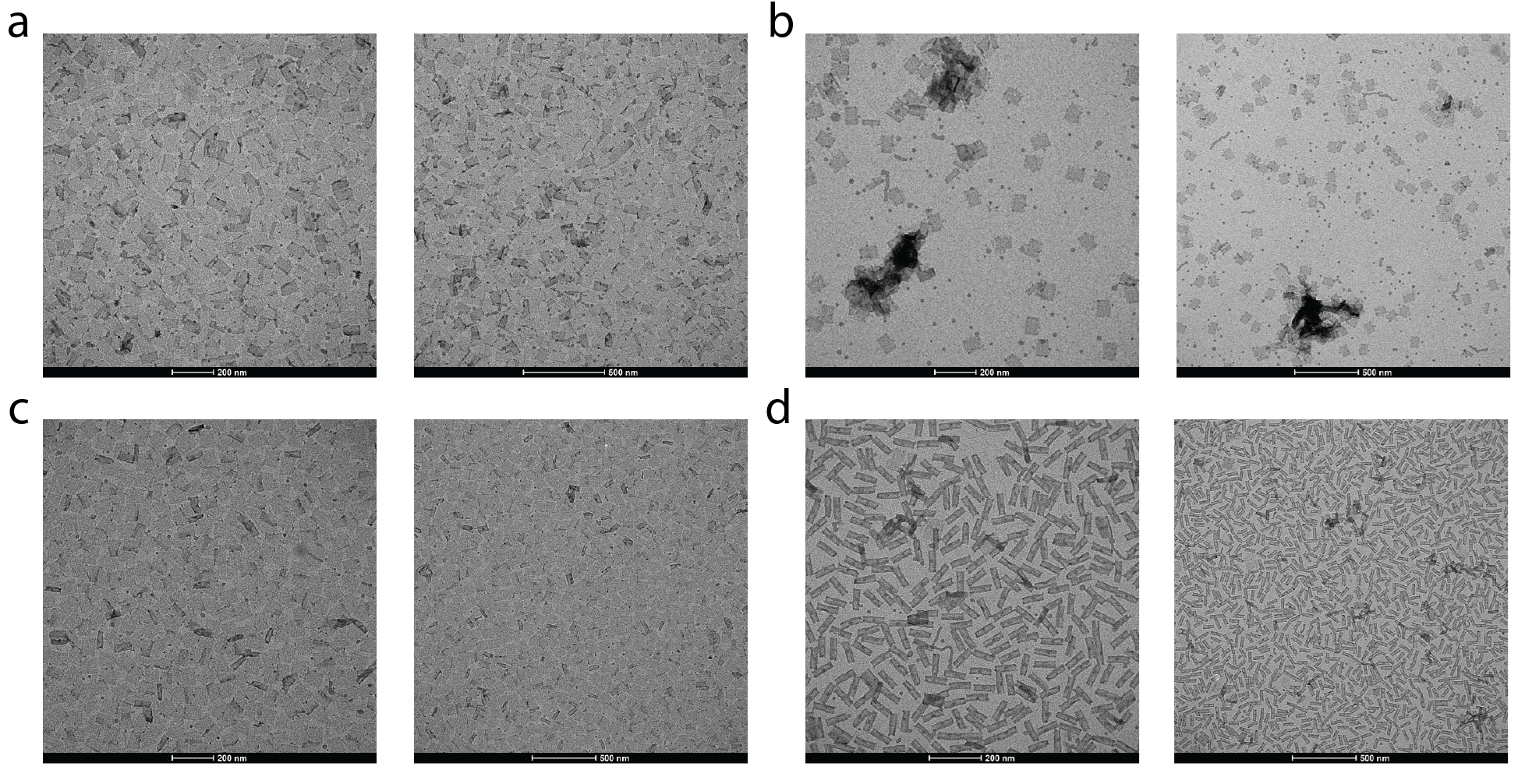}
    \caption{Negative staining TEM images of twist corrected DNA origami rectangles folded in 1x TAE-$\mathrm{Na^{+}}$ buffer with $(a)$ no overhang and no locking strand, $(b)$ no overhangs with locking strands, $(c)$ 20nt overhang extensions with no locking strand, $(d)$ 20nt overhang extensions with locking strand.}    
    \label{fig:TEMSodium}
    \end{figure*}

\begin{figure*}[!ht]
     \centering
     \includegraphics[width=0.9\textwidth]{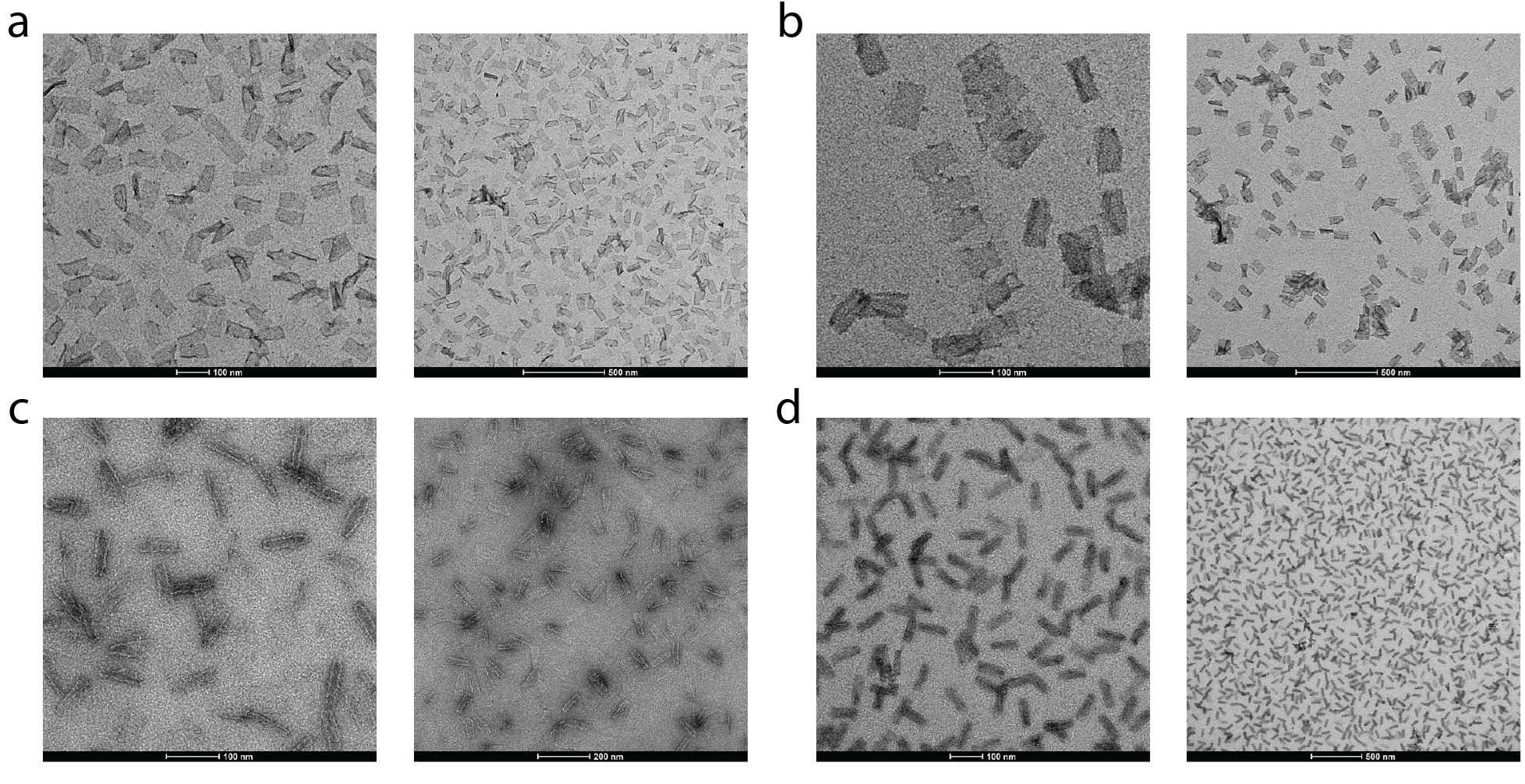}
    \caption{Negative staining TEM images of protein-conjugated rectangle DNA origami with $(a)$ no overhang and no locking strand, $(b)$ no overhangs with locking strands, $(c)$ 20nt overhang extensions with no locking strand, $(d)$ 20nt overhang extensions with locking strand. Biotinylated strands and the protein are added under all conditions.}    
    \label{fig:TEMprotein}
    \end{figure*}

\begin{figure*}[!ht]
     \centering
     \includegraphics[width=0.9\textwidth]{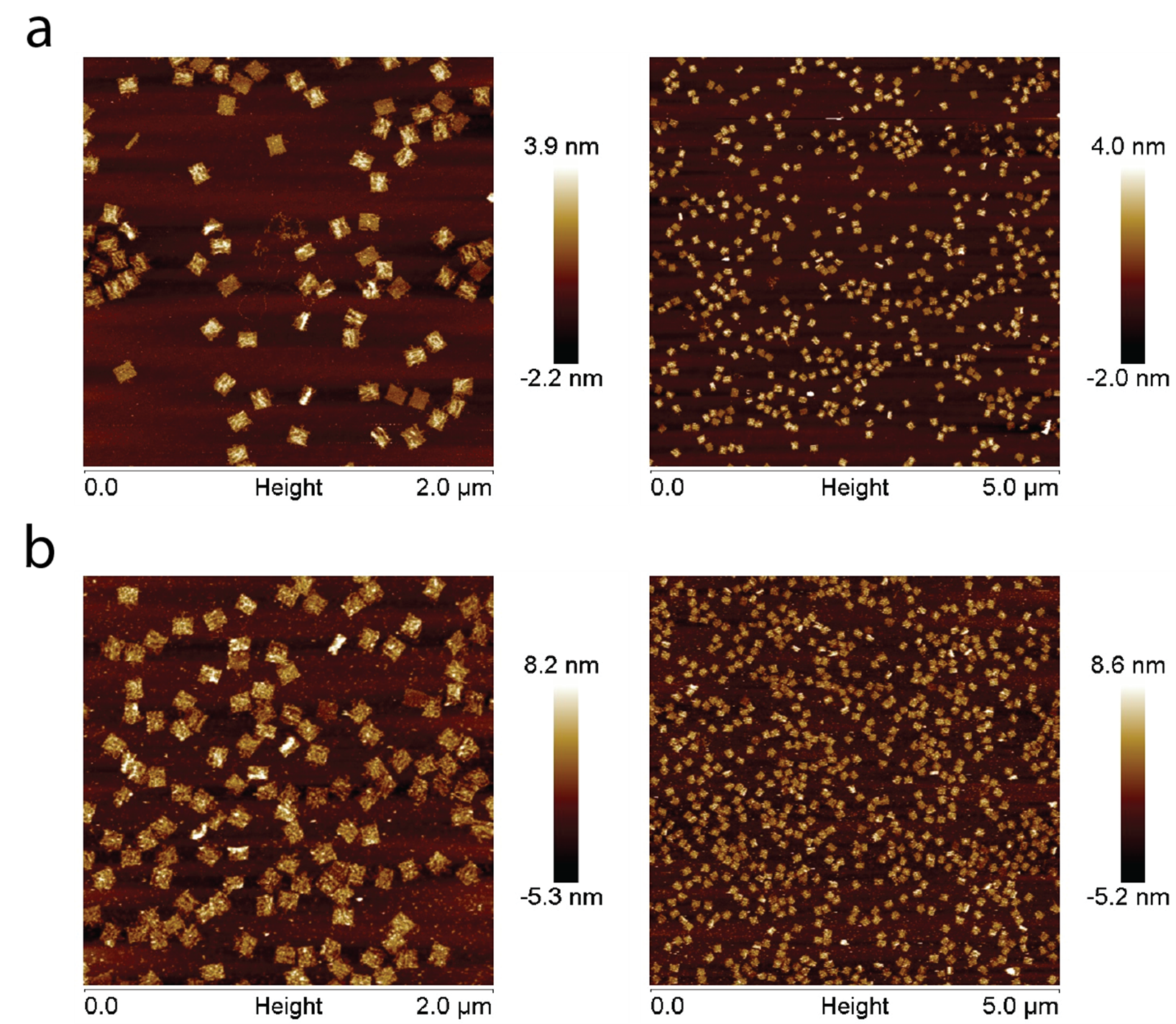}
    \caption{AFM images of protein-conjugated rectangle origami experiment. $(a)$ 20 nt overhang extensions with only biotinylated strands. $(b)$ 20 nt overhang extensions with both biotinylated strands and protein added. An increase in the height of the rectangle origami, indicates that proteins are successfully conjugated. While a curvature is expected, due to the intrinsic properties of AFM, the origami sheets appear mostly flat.}    
    \label{fig:AFMprotein}
    \end{figure*}

\begin{figure*}[!ht]
     \centering
     \includegraphics[width=0.9\textwidth]{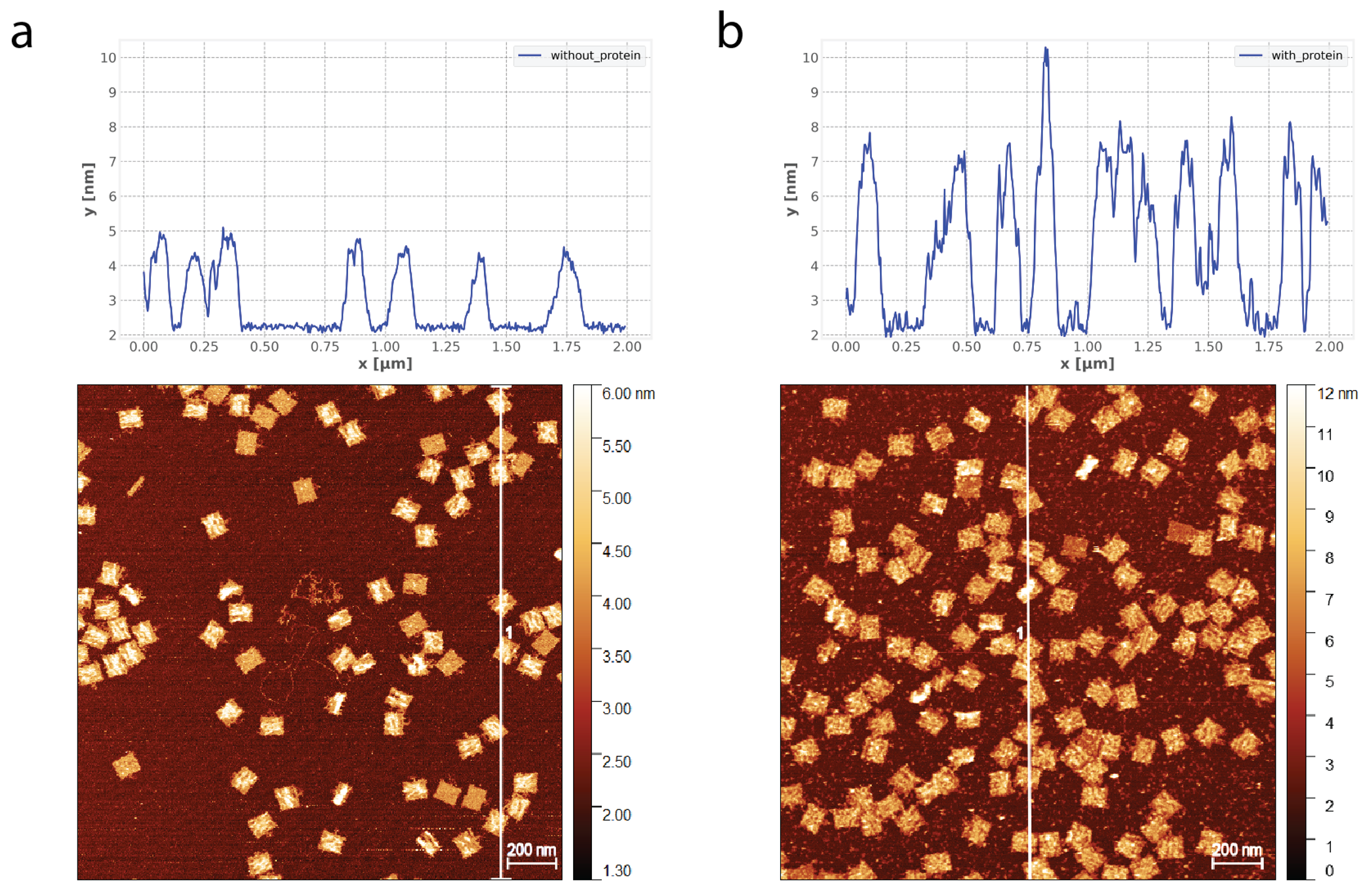}
    \caption{AFM height profiles of protein-conjugated rectangle origamis. $(a)$ 20 nt overhang extensions with only biotinylated strands. $(b)$ 20 nt overhang extensions with both biotinylated strands and protein added. The height profiles show the increased height resulting from the conjugation of the proteins.}    
    \label{fig:AFMproteinheight}
    \end{figure*}

\section{Umbrella Sampling}
\subsection{Background}
Umbrella sampling is an enhanced sampling molecular dynamics technique utilized to calculate free energy profiles as a function of a chosen order parameter. An order parameter is chosen for its ability to represent a phenomenon of interest. The umbrella sampling technique uses multiple simulation replicas called windows, where each window samples a different subset of your order parameter. By applying a biasing potential, called the umbrella potential, to each simulation window we are able to explicitly sample the entire range of values along the axis of our order parameter. After all simulation windows have run, we unbias the simulation results using the Weighted Histogram Analysis Method (WHAM)\cite{torrie1974monte, torrie1977nonphysical, wham}. Using multiple simulation replicas to sample different order parameter values relevant to our phenomenon of interest decreases the time required for our free energy profile to converge.

\subsection{Order Parameter}
  \begin{figure}
     \centering
     \includegraphics[width=0.425\textwidth]{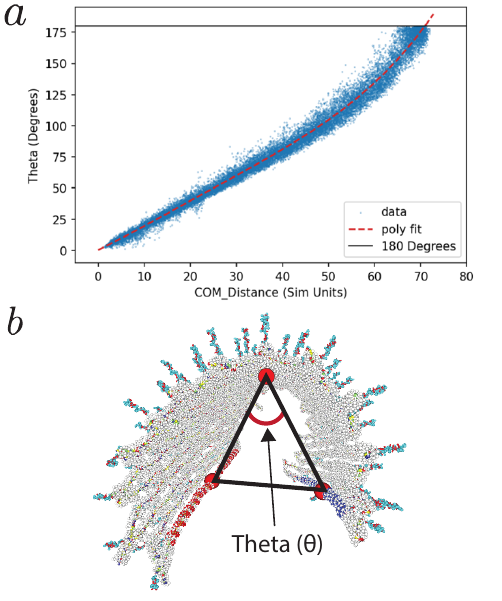}
    \caption{$(a)$ The angle theta as a function of the end-to-end distance was fit using polynomial regression. Theta was calculated by measuring the three distances shown in b) and using the law of cosines relationship to get theta. The fit shows that as the value of $R_{\rm ee}$ increases the angle theta also increases. As a theta value of 180$^{\circ}$ indicates a flat structure, as $R_{\rm ee}$ increases the curvature decreases.}    \label{fig:a_r}
    \end{figure}
To quantify the curvature of the rectangular structure the order parameter we chose was the distance between the center of mass of the scaffold nucleotides along the long edge of the rectangular structure, called the end to end distance ($R_{\rm ee}$). We excluded nucleotides at the corners of the rectangular structure from our order parameter as they experience large magnitude random fluctuations and transient hydrogen bond fraying. 

The $R_{\rm ee}$ was chosen for a few reasons. First, the relationship between the distance of the two ends and the angle theta (shown in Fig S1) provides an metric to quantify the relative curvature of the rectangular origami. Additionally, as $R_{\rm ee}$ is a 1D parameter, we get faster convergence of our free-energy landscapes as sampling all possible values of a single degree of freedom can be done quickly. Finally, as our chosen umbrella potential was a COM harmonic biasing potential, choosing an OP based on the distance between two centers of masses enables us to directly bias our structures to sample our OP.

The value of $R_{\rm ee}$ was sampled from 0.62 nm to 62 nm. A hundred windows were used where each window sampled OP values biased to fluctuate around specific  $R_{\rm ee}$ values. Equally spaced increments of 0.62 nm along the 1D axis of the OP were used. The extrema values of $R_{\rm ee}$ was chosen by preforming pulling simulations on the 0 overhang structure using an external attraction plane potential. 

The attraction plane potential was applied to all nucleotides in the system, pulling them towards a plane on the x-y axis with a stiffness of 0.1 pN/nm per nucleotide. oxView was then used to identify the nucleotide IDs along each of the rectangles long edges and oxDNA-analysis-tools was used to calculate the mean distance between the COM of the two edges. An extra 2 nm was added to the mean distance and used as the max $R_{\rm ee}$ distance (62 nm) to sample states beyond the most stable flat structure.

\begin{figure}
\centering
 \includegraphics[width=0.425\textwidth]{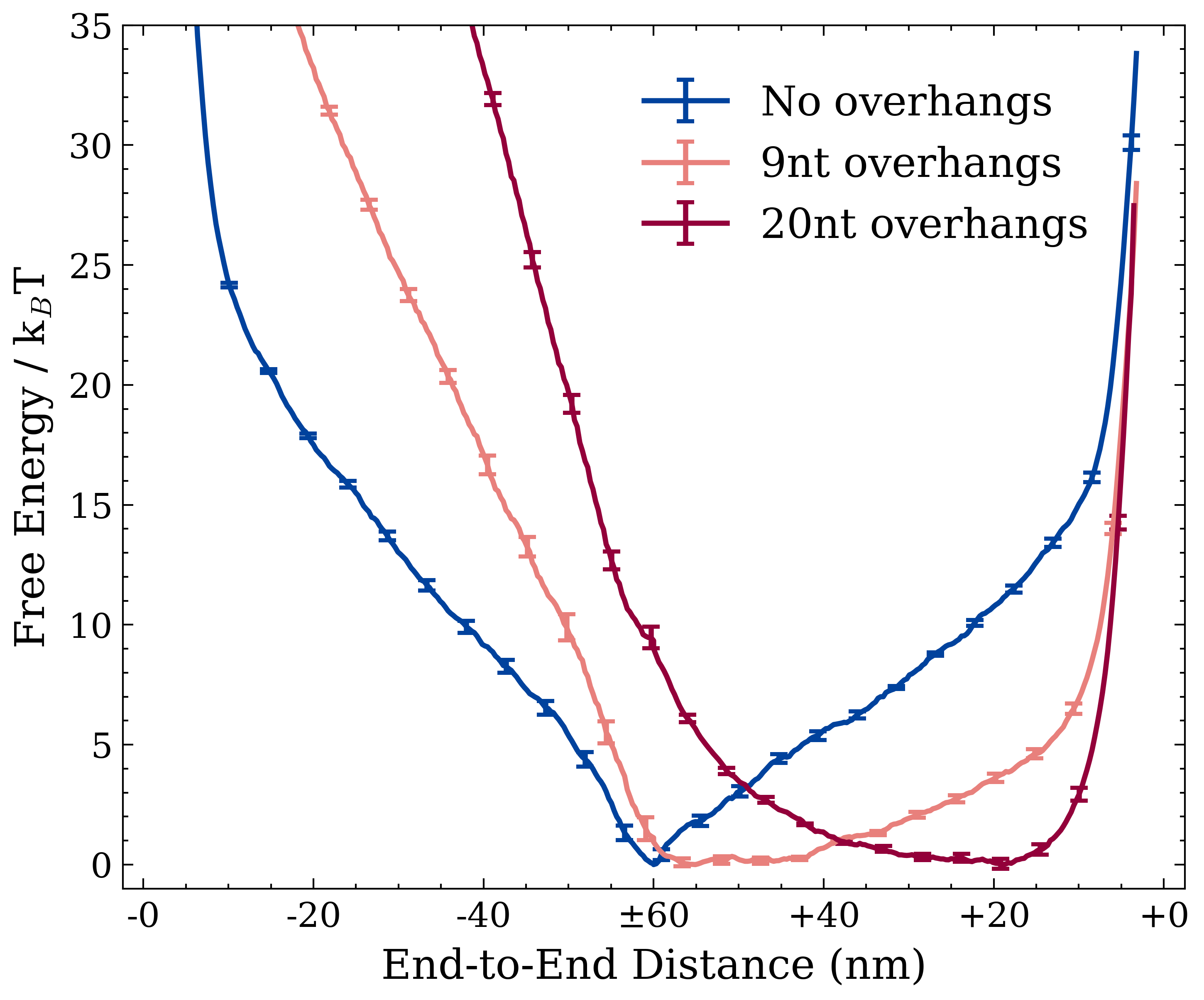}
\caption{Umbrella profiles of structures with overhangs extensions on both the interior and exterior of the curvature. The negative End-to-End distance values indicate structures where the overhangs are on the inside of the structure, while positive values are states with the overhangs on the outside.}    \label{fig:inverted}
\end{figure}
\nopagebreak
    \begin{figure}
    \centering
     \includegraphics[width=0.45\textwidth]{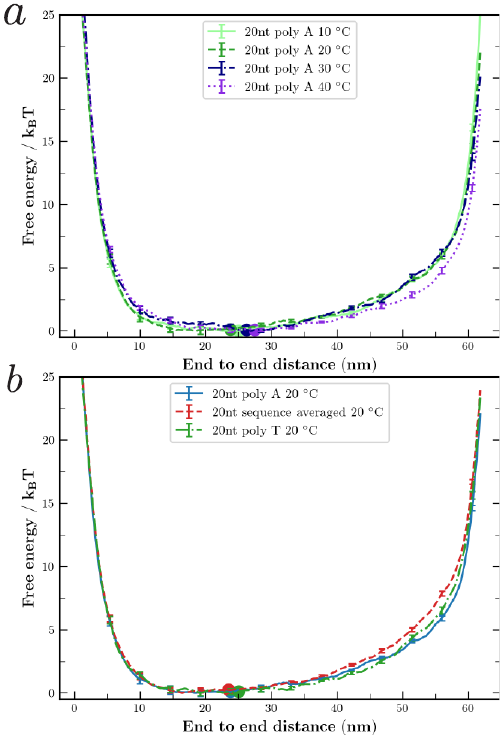}
    \caption{Effect of overhang nucleotide sequence. The difference between the 20nt poly A and poly T simulations show no significant difference. When the temperature is varied for the poly A structure no large difference is seen for the weighted average.}    \label{fig:a_r}
    \end{figure}
    \nopagebreak
\nopagebreak
    \begin{figure}
    \centering
     \includegraphics[width=0.50\textwidth]{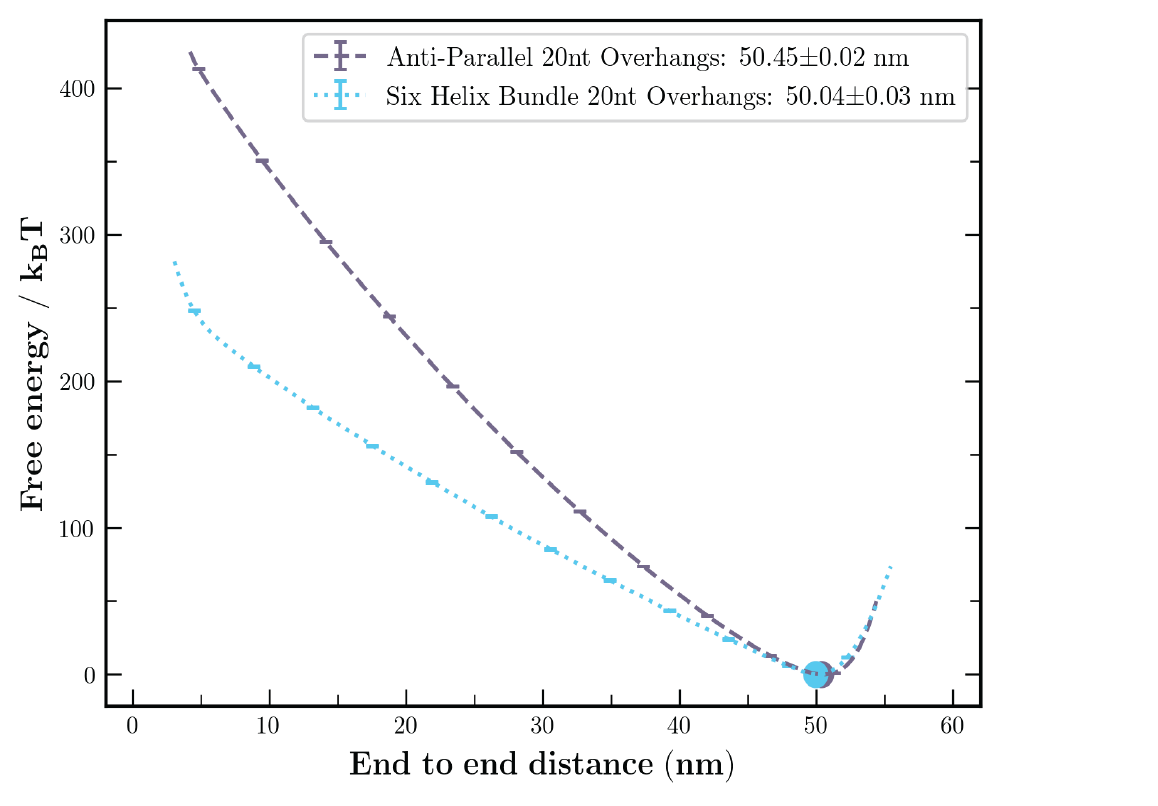}
    \caption{Full free energy profiles of 3D structures as a function of end to end distance. The plot shows that while both structure require enormous energies to curve, the antiparallel double layered sheet shows a greater resistance to bending.}    \label{fig:a_r}
    \end{figure}
    \nopagebreak
\begin{figure*}
\centering
\includegraphics[width=0.80\textwidth]{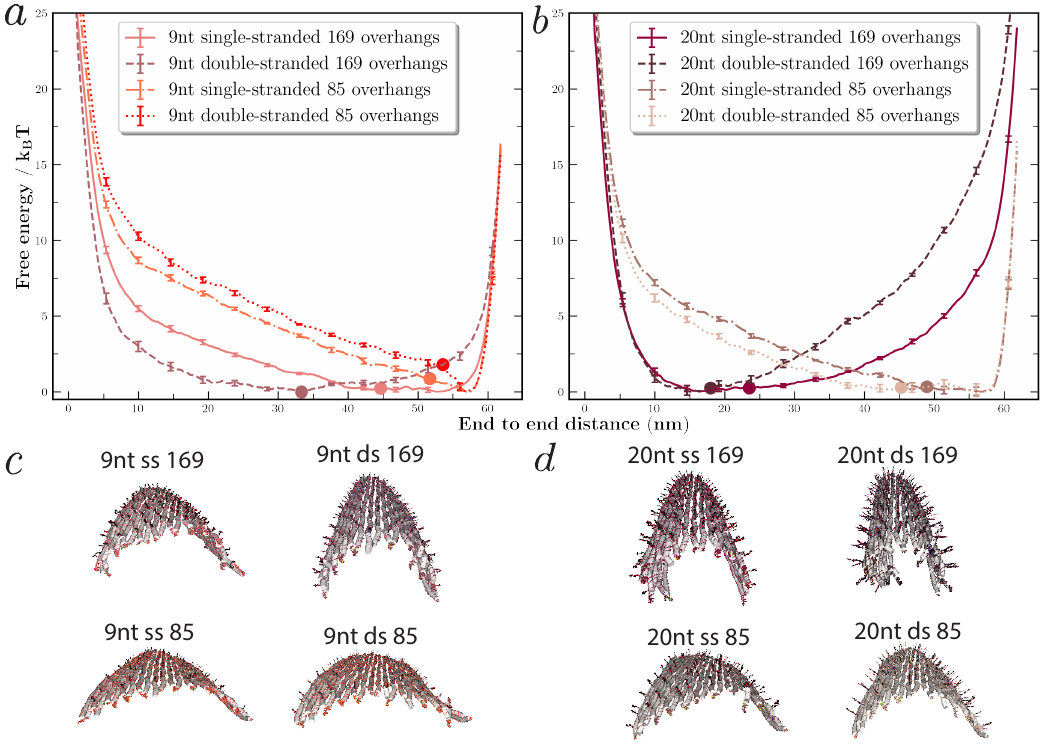}
\caption{Effect of different overhang conditions. Free energy profile of $(a)$ $9$nt  and $(b)$ $20$nt modified overhang systems. When the complementary strand of the overhang extension is added to create double-stranded overhangs, the curvature of both the $9$nt and $20$nt structures increases. Inversely, if the number of single-stranded extensions is reduced to $85$ overhangs, the curvature of the structures decrease. Mean structures of $(c)$ $9$nt and $(d)$ $20$nt overhang conditions illustrate these differences.}    \label{fig:duplex}
\end{figure*}
\nopagebreak
\begin{figure*}[!ht]
\centering
\includegraphics[width=0.80\textwidth]{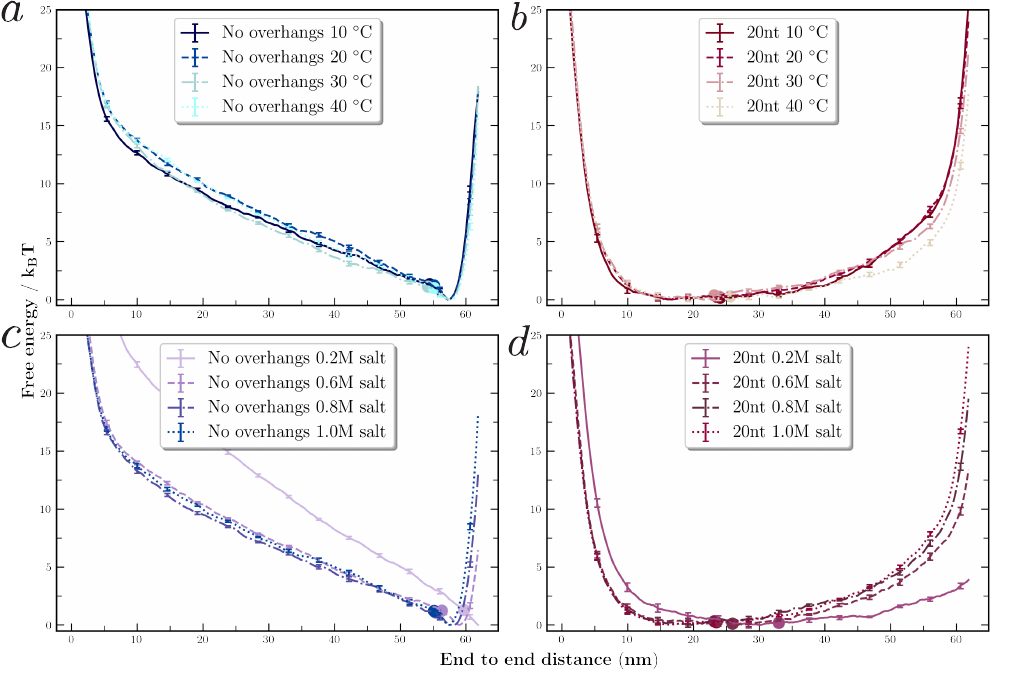}
\caption{Effect of different temperatures and salt concentrations. Free energy profiles of $(a, c)$ no overhangs and $(b, d)$ $20$nt overhangs at various temperatures and salt concentrations. For both no overhang and $20$nt structures, when the temperature is increased the curvature is negligibly effected. When the salt concentration is decreased, the curvature of the structure decreases due to the increase in electrostatic repulsion between charged backbone sites in the rectangular tile.}    \label{fig:temp_mol_3}
\end{figure*}
\nopagebreak
\begin{figure*}
\centering
\includegraphics[width=0.95\textwidth]{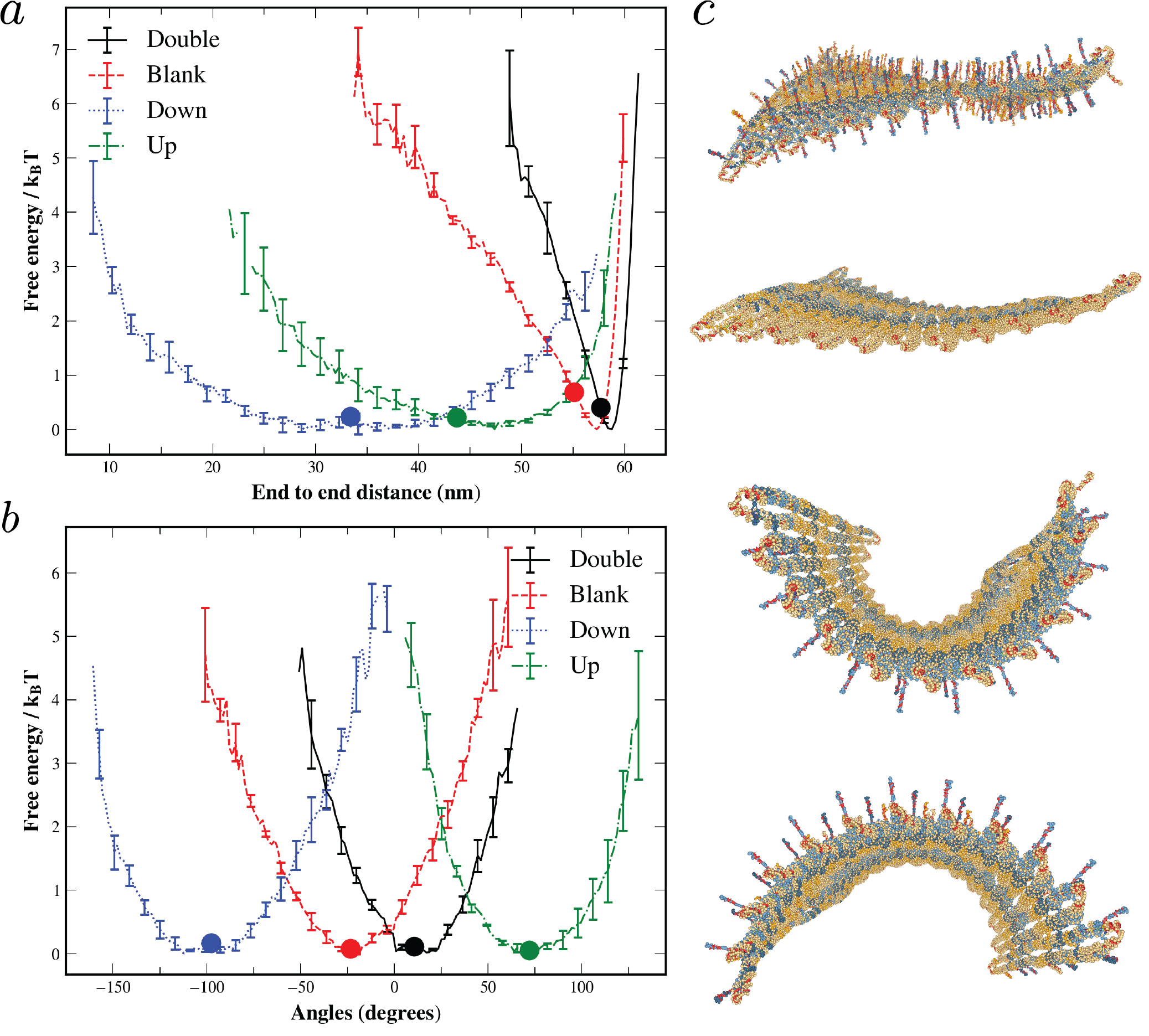}
\caption{Effect of having overhangs on both sides. Free energy profile of as a function of $(a)$ End to end distance and $(b)$ angle of double sided overhang systems from "normal" brute force simulations. When there are overhang extensions on both sides of the rectangular origami, the origami remains flat. For the blank system, the slight bias on the curvature leads to an average negative curvature. This bias is seen in the down and up systems, where the down system show higher curvature due to the bias and conversely the up system has less curvature.}    \label{fig:duplex}
\end{figure*}
\nopagebreak
\subsection{Running Umbrella Sampling}
\nopagebreak

We have provided an interactive Jupyter notebook to automate the majority of the tasks required to run umbrella sampling which can be found here: \url{https://github.com/mlsample/hairygami_umbrella_sampling}
 As a prerequisite, oxDNA python environment (oxpy) must be installed. It can be obtained though compiling oxDNA with python support (see \url{https://lorenzo-rovigatti.github.io/oxDNA/install.html}).  The parameters used to run the umbrella simulations can be found in the tutorial notebook. 
\nopagebreak

 To ensure proper equilibration of starting conformations was preformed, we trimmed off the first 1 million, 2 million, and 3 million productions steps of each simulation window. The trimmed profiles look identical to the profile with all the data indicating that the starting conformation of the umbrella sampling simulations were equilibrated properly. To check convergence we separated the first and last 10 million steps, as well as plotting free energy profiles of data points obtained from 10, 12, 14, 16, and 18 million steps. As the free energy profiles of both the first and last 10 million steps are nearly identical, it can be said that the umbrella sampling simulations have converged. Similarly, as we increased the number of data points the the free energy profile negligibly changed as the umbrella windows were run longer.

\clearpage

 \onecolumngrid
 
\begin{table}
\scalebox{1.0}{
\begin{tabular}{ |c||c | c| }
\hline
System &\makecell{End-to-End\\Distance}& \makecell{Standard Error\\ of the Mean}\\
\hline
$0$ Hangs $0.2$ M Salt&$59.77$&$0.10$\\
$0$ Hangs $0.6$ M Salt&$56.33$&$0.13$\\
$0$ Hangs $0.8$ M Salt&$55.37$&$0.12$\\
$0$ Hangs $10$ $^{\circ}$C&$54.50$&$0.13$\\
$0$ Hangs $20$ $^{\circ}$C &$55.07$ &$0.11$\\
$0$ Hangs $30$ $^{\circ}$C&$54.07$ &$0.14$\\
$0$ Hangs $40$ $^{\circ}$C&$55.04$ &$0.13$\\
$1$nt $20$ $^{\circ}$C&$53.03$ &$0.09$\\
$3$nt $20$ $^{\circ}$C&$50.59$ &$0.12$\\
$9$nt $10$ $^{\circ}$C&$42.56$ &$0.14$\\
$9$nt $20$ $^{\circ}$C&$44.74$ &$0.10$\\
$9$nt $40$ $^{\circ}$C&$45.71$ &$0.13$\\
$9$nt Half Density&$51.75$ &$0.16$\\
$9$nt Duplex $10$ $^{\circ}$C&$35.30$ &$0.14$\\
$9$nt Duplex $20$ $^{\circ}$C&$33.20$ &$0.14$\\
$9$nt Duplex $40$ $^{\circ}$C&$39.75$ &$0.16$\\
$9$nt Duplex Half Density&$53.70$ &$0.15$\\
$9$nt Overhang Clashing Off&$44.60$ &$0.16$\\
$9$nt Tile Clashing Off&$53.62$ &$0.10$\\
$9$nt Tile and Overhang Clashing Off&$53.36$ &$0.11$\\
$9$nt Poly T $10$ $^{\circ}$C&$43.98$ &$0.12$\\
$9$nt Poly T $40$ $^{\circ}$C&$48.52$ &$0.11$\\
$20$nt $0.2$ M Salt&$32.88$ &$0.18$\\
$20$nt $0.6$ M Salt&$25.81$ &$0.14$\\
$20$nt $0.8$ M Salt&$23.14$ &$0.15$\\
$20$nt $10$ $^{\circ}$C&$24.08$ &$0.21$\\
$20$nt $20$ $^{\circ}$C&$23.48$ &$0.12$\\
$20$nt $30$ $^{\circ}$C&$23.03$ &$0.18$\\
$20$nt $40$ $^{\circ}$C&$25.76$ &$0.14$\\
$20$nt Half Density&$48.87$ &$0.18$\\
$20$nt Duplex&$18.04$ &$0.17$\\
$20$nt Duplex Half Density&$45.23$ &$0.24$\\
$20$nt Overhang Clashing Off $10$ $^{\circ}$C&$36.15$ &$0.17$\\
$20$nt Overhang Clashing Off $20$ $^{\circ}$&$37.66$ &$0.17$\\
$20$nt Overhang Clashing Off $40$ $^{\circ}$C&$41.48$ &$0.17$\\
$20$nt Tile Clashing Off $10$ $^{\circ}$C&$53.08$ &$0.13$\\
$20$nt Tile Clashing Off $20$ $^{\circ}$C&$54.05$ &$0.13$\\
$20$nt Tile Clashing Off $40$ $^{\circ}$C&$55.26$ &$0.15$\\
$20$nt Tile and Overhang Clashing Off&$54.78$ &$0.14$\\
$20$nt Duplex $10$ $^{\circ}$C&$16.69$ &$0.16$\\
$20$nt Duplex $40$ $^{\circ}$C&$19.24$ &$0.16$\\
$20$nt Poly T $10$ $^{\circ}$C&$23.65$ &$0.22$\\
$20$nt Poly T $20$ $^{\circ}$C&$25.16$ &$0.16$\\
$20$nt Poly T $30$ $^{\circ}$C&$23.32$ &$0.18$\\
$20$nt Poly T $40$ $^{\circ}$C&$26.84$ &$0.15$\\
$20$nt Poly A $10$ $^{\circ}$C & $25.10$ & $0.14$\\
$20$nt Poly A $20$ $^{\circ}$C&$23.94$ &$0.19$\\
$20$nt Poly A $30$ $^{\circ}$C&$26.33$ &$0.14$\\
$20$nt Poly A $40$ $^{\circ}$C&$27.45$ &$0.12$\\
$20$nt Double-Layer Anti-parallel $20$ &$50.49$ &$0.02$\\
$20$nt Six Helix Bundle $20$ &$50.04$ &$0.03$\\

 \hline
\end{tabular}
}
\label{fig:temp_mol}
\caption{Weighted average end-to-end distance of all simulated umbrella conditions}
\end{table}

\begin{figure}
     \centering
     \includegraphics[width=0.95\textwidth]{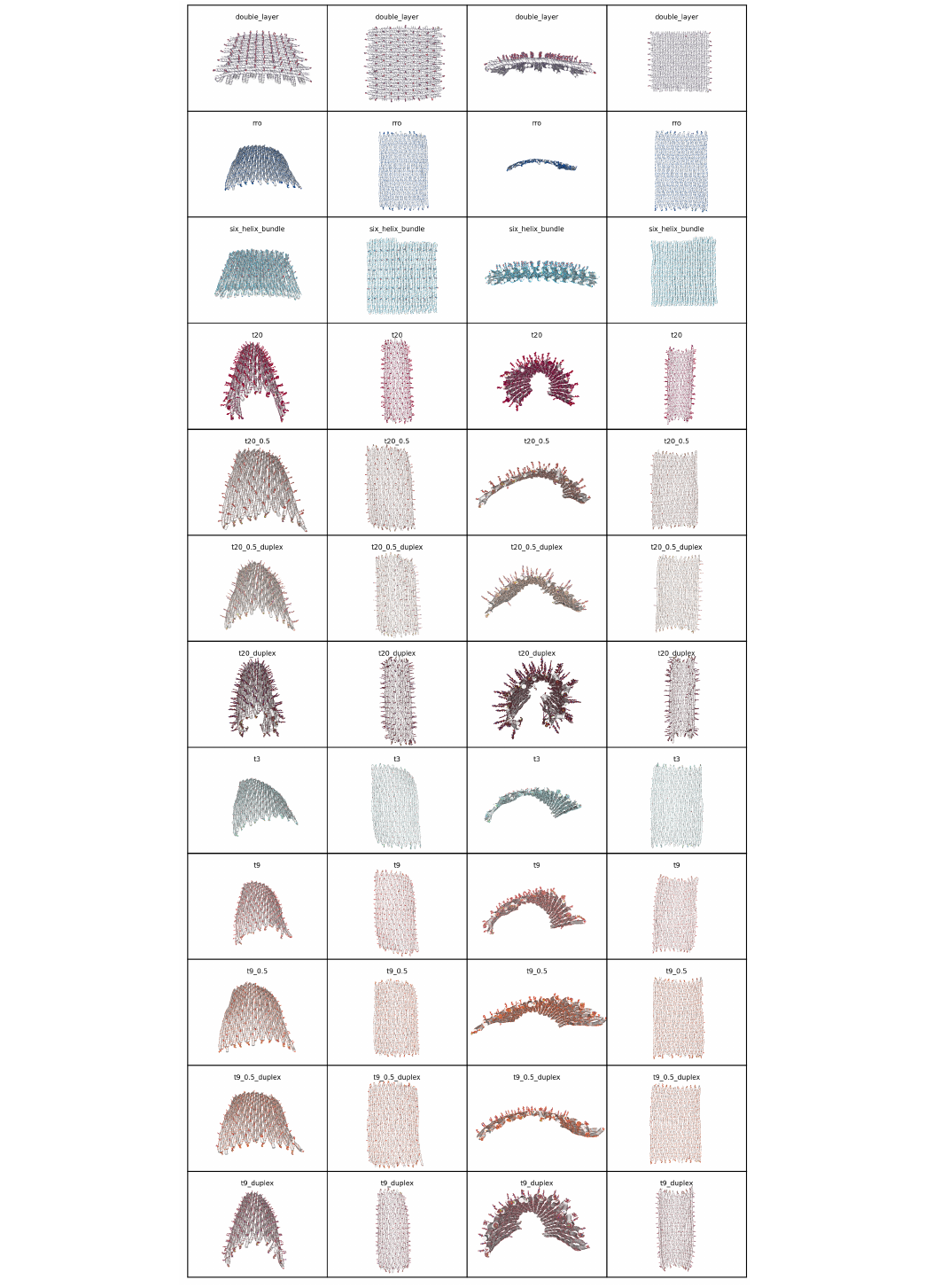}
    \caption{Means of simulated structures}    
    \label{fig:collage}
    \end{figure}
\clearpage




\begin{longtblr}[caption = {Staples for rectangular origami without poly-T overhangs}]
{ 
  colspec = {|XX[4]|},
  hlines,
} 
Core S1& \seqsplit{ATACCGATTTTCCAGACGTTAGTAACCAGTACAAACTAC}\\
Core S2& \seqsplit{TTCACAAAATGCCCCCTGCCTATTGGATAAGTGCCGTCG}\\
Core S3& \seqsplit{GGTTGAGGTGAAACATGAAAGTAAGGATTAGCGGGGT}\\
Core S4& \seqsplit{TGTCGTCAGTTGCGCCGACAATATTCGGTC}\\
Core S5& \seqsplit{TTCTGTATGAGGTGAATTTCTTAAGGCCGCTTTTGCGGGAAAGAA}\\
Core S6& \seqsplit{ACAGTTTCTTTAATTGTATCGGTTGCGAAAGA}\\
Core S7& \seqsplit{AACTAAAAATCTCCAAAAAAAAGGCTACAGAGGCTTTCATTAAA}\\
Core S8& \seqsplit{ACGTTGAAGGAATTGCGAATAATATTGATGAT}\\
Core S9& \seqsplit{ACAGGAGTGCGTCATACATGGCTTATTTTTTC}\\
Core S10& \seqsplit{ACAGTTACAAATAAATCCTCATCTCCCTCA}\\
Core S11& \seqsplit{GCTGAGGCAGCGATTATACCAAATCGCCTG}\\
Core S12& \seqsplit{TTGACCCCCTTGCAGGGAGTTAAAACAGCTTG}\\
Core S13& \seqsplit{AGAGGCAATCGTCACCCTCAGCATATCAGC}\\
Core S14& \seqsplit{CAGCATCGCGAAGGCACCAACCTAGCAGACGG}\\
Core S15& \seqsplit{TGCCACTAGAACGAGGGTAGCAACGGCTCCAA}\\
Core S16& \seqsplit{GAAGTTTCGAGGACTAAAGACTTTTATTAGCG}\\
Core S17& \seqsplit{TTTGCCATTCGGTCATAGCCCCCTTTCATGAG}\\
Core S18& \seqsplit{CGGCATTTCTTTTCATAATCAAATTACCGT}\\
Core S19& \seqsplit{CCTTTAGCCCACCACCGGAACCGCTAAAGCCA}\\
Core S20& \seqsplit{GAGCCGCCAGTAGCGACAGAATAATTATTC}\\
Core S21& \seqsplit{ACCGTAATCACCCTCAGAACCGCCCCTTGATA}\\
Core S22& \seqsplit{GAAACGTCCCCTCAGAGCCGCCAGACAGGA}\\
Core S23& \seqsplit{ATAAATTTTGCCCTGACGAGAATGGTTTAA}\\
Core S24& \seqsplit{AATAAGGCGTGTCGAAATCCGCGAACTCATCT}\\
Core S25& \seqsplit{CAAATCAATAGCCGGAACGAGGCAAACGAA}\\
Core S26& \seqsplit{TCAATCATACAAGAACCGGATATTATACCAGT}\\
Core S27& \seqsplit{TAATCTTGAAGGGAACCGAACTGAATACGTAA}\\
Core S28& \seqsplit{GGCGCATAGACAGATGAACGGTGTCAGCGCCA}\\
Core S29& \seqsplit{AAGACAAAAATTCATATGGTTTACACAGACCA}\\
Core S30& \seqsplit{CAATAGAAAGGGCGACATTCAACTTTTCAT}\\
Core S31& \seqsplit{ACACCACGGGTAAATATTGACGGACAAGTTTG}\\
Core S32& \seqsplit{ATTAAAGCAACATATAAAAGAAGTAAGCAG}\\
Core S33& \seqsplit{AAAGGTGGGTGAATTATCACCGTCATAGCAGC}\\
Core S34& \seqsplit{AGTATGTTTGGGAATTAGAGCCAAAGGCCG}\\
Core S35& \seqsplit{TTTCAACAACCCTCGTTTACCATTTGCAAA}\\
Core S36& \seqsplit{ACTATCATTTTAATCATTGTGAATATTCAGTG}\\
Core S37& \seqsplit{CAAAAGGAAAGAACTGGCTCATTCATTACC}\\
Core S38& \seqsplit{CAGGACGTCAACTAATGCAGATACTACTGCGG}\\
Core S39& \seqsplit{ACCACATTTGGGAAGAAAAATCTAATCAAGAG}\\
Core S40& \seqsplit{TAGAAAGACTAACGGAACAACATTATAATAAG}\\
Core S41& \seqsplit{AGCAAGAAAATTGAGTTAAGCCCAATTACAGG}\\
Core S42& \seqsplit{CCCACAAGACAATGAAATAGCAATCACAAT}\\
Core S43& \seqsplit{GGGTAATTGCCCTTTTTAAGAAAAACGCAAAG}\\
Core S44& \seqsplit{ATAGCCGTGAACACCCTGAACACAGTTACA}\\
Core S45& \seqsplit{GAATTAACAACAAAGTTACCAGAAACATACAT}\\
Core S46& \seqsplit{AACATAAACAATAATAACGGAATATTACGC}\\
Core S47& \seqsplit{AGAAGTTGCGTTTTAATTCGAGAACAGGTC}\\
Core S48& \seqsplit{CAAATATCTTGCCAGAGGGGGTAAAGAGCAAC}\\
Core S49& \seqsplit{TGTTTAGAATTAAGAGGAAGCCCGGCTCCTTT}\\
Core S50& \seqsplit{TCAAAAAGCTGGATAGCGTCCAAATAACGC}\\
Core S51& \seqsplit{AATCGTCAAGTCAGAAGCAAAGCGTGGCTTAG}\\
Core S52& \seqsplit{ACTATTATTAAATATTCATTGAATTTAGGAAT}\\
Core S53& \seqsplit{AATGCTTTCAAAAATCAGGTCTGTAGCTCA}\\
Core S54& \seqsplit{ACCATAAATAAACAGTTCAGAAAACAGCTACA}\\
Core S55& \seqsplit{ATTTTATCGTTGCTATTTTGCACCCGAGAATG}\\
Core S56& \seqsplit{CAAGATTACTGAATCTTACCAACGAGATAA}\\
Core S57& \seqsplit{GCGTCTTTCGGGAGGTTTTGAAGCTAGAAACC}\\
Core S58& \seqsplit{CCGACTTGCCAGAGCCTAATTTGCAAGTCAGA}\\
Core S59& \seqsplit{AAATAAACGCGAGGCGTTTTAGTATCATTC}\\
Core S60& \seqsplit{TCTAAGAACAGCCATATTATTTATAGACGGGA}\\
Core S61& \seqsplit{AAATAAGAAGATATAGAAGGCTTAACCGCACT}\\
Core S62& \seqsplit{AGCAAATCAACGATTTTTTGTTTAGAGAAT}\\
Core S63& \seqsplit{AGGATTATAGCTATATTTTCATCTACTAAT}\\
Core S64& \seqsplit{AACCTGTTGAGAGTACCTTTAATTAAAGACTT}\\
Core S65& \seqsplit{TGATAAGAAGATACATTTCGCAAAAATCATAC}\\
Core S66& \seqsplit{TGACCATTGGTCATTTTTGCGGAGATTGCA}\\
Core S67& \seqsplit{AGCTTAATTCTGCGAACGAGTAGAATTAAGCA}\\
Core S68& \seqsplit{TTCCCAATTGCTGAATATAATGCTTTACCCTG}\\
Core S69& \seqsplit{ACATGTTGTTTCATTCCATATAAATCGGTT}\\
Core S70& \seqsplit{GTCTGGAATTAAATATGCAACTAAAAATAATA}\\
Core S71& \seqsplit{TCCCATCCAAGTCCTGAACAAGAAAGTACGGT}\\
Core S72& \seqsplit{CAATAGATTAATTTACGAGCATGCTTAAAT}\\
Core S73& \seqsplit{AATCAATATGCAGAACGCGCCTGTAGTATCAT}\\
Core S74& \seqsplit{TCAGCTAAATCGGCTGTCTTTCCTCGAACCTC}\\
Core S75& \seqsplit{CAAGAACGACGACGACAATAAATATAAAGC}\\
Core S76& \seqsplit{TCTGTCCAGGGTATTAAACCAAGTTCCGGTAT}\\
Core S77& \seqsplit{CATCGAGAAGTACCGACAAAAGGTTTGAGAAT}\\
Core S78& \seqsplit{GAATATAAACAAGCAAGCCGTTTAATAGCA}\\
Core S79& \seqsplit{AGTAGTAGGTGAGAAAGGCCGGACCGTTCT}\\
Core S80& \seqsplit{TTCAAAAGGCATTAACATCCAATATGGTCAAT}\\
Core S81& \seqsplit{AGGCAAGGGCCTGAGTAATGTGTAGGGTAGCT}\\
Core S82& \seqsplit{AATGCAATCAAAGAATTAGCAAATTTAGTT}\\
Core S83& \seqsplit{ATAAAGCCTTTTAGAACCCTCATATATCAGGT}\\
Core S84& \seqsplit{ATAAAAATTCAGAGCATAAAGCTAACAGTTGA}\\
Core S85& \seqsplit{GTACCAAGAAGCCTTTATTTCACAAGAGAA}\\
Core S86& \seqsplit{TTGCGGGAAAACATTATGACCCTGCGGAATCA}\\
Core S87& \seqsplit{TAATTACTAAATAAGAATAAACACTAATACTT}\\
Core S88& \seqsplit{AAGGCGTTAGAAAAAGCCTGTTTTTATCAA}\\
Core S89& \seqsplit{ATGCGTTAGAAATACCGACCGTGTTAGATTAA}\\
Core S90& \seqsplit{AATGGTTTTACAAATTCTTACCAGCAACATGT}\\
Core S91& \seqsplit{CAACGCTATTTCATCTTCTGACATTTATCA}\\
Core S92& \seqsplit{TTTAGTTACAACAGTAGGGCTTAAAAAGTAAT}\\
Core S93& \seqsplit{CGCCATATGCGAGAAAACTTTTTCCTTTTTAA}\\
Core S94& \seqsplit{CAAAGAACTTAACAACGCCAACAAATAAGA}\\
Core S95& \seqsplit{AGCTGATCATTAAATTTTTGTTCATCAAAA}\\
Core S96& \seqsplit{AAAATTCGAAATTAATGCCGGAGAGGTAAAGA}\\
Core S97& \seqsplit{AATATTTAGAGATCTACAAAGGCTATTTTA}\\
Core S98& \seqsplit{CATTGCCTAAAACAGGAAGATTGTGAGTAACA}\\
Core S99& \seqsplit{AAGCCCCAGAGAGTCTGGAGCAAAACGCAAGG}\\
Core S100& \seqsplit{TCGATGATGTACCCCGGTTGATAAACGGCGGATTGACGCGCATC}\\
Core S101& \seqsplit{CAATCATAACGGTAATCGTAAAACCCTTAGAA}\\
Core S102& \seqsplit{TCCTTGAACTATTAATTAATTTTCTAGCATGT}\\
Core S103& \seqsplit{AATCGTCGAACATAGCGATAGCTGATAAAT}\\
Core S104& \seqsplit{ATATATGTGAAGAGTCAATAGTGACTAAATTT}\\
Core S105& \seqsplit{AAATCATTAATGGAAACAGTACTGATTGCT}\\
Core S106& \seqsplit{ACCTTTTTAGGTCTGAGAGACTACAAATATAT}\\
Core S107& \seqsplit{CCTCCGGCACATTTAACAATTTCACGCAGAGGCGAATTATTCCTG}\\
Core S108& \seqsplit{AATTAATTTTAGGTTGGGTTATACGCAAGA}\\
Core S109& \seqsplit{ATAATTCTGGTGCCGGAAACCAACTGTTGG}\\
Core S110& \seqsplit{ACCGCTTCGCGTCTGGCCTTCCTGATTTTGTT}\\
Core S111& \seqsplit{CAGGAAGAACATTAAATGTGAGCATAAGCA}\\
Core S112& \seqsplit{ACCCGTCGGGGGACGACGACAGTATGTGCTGC}\\
Core S113& \seqsplit{CAGTTTGAGATTCTCCGTGGGAACAATCAGAA}\\
Core S114& \seqsplit{GTAGATGGCGTAATGGGATAGGTCAACGTCAG}\\
Core S115& \seqsplit{ATGAATATGTAGATTTTCAGGTTTACGTTGGT}\\
Core S116& \seqsplit{GAAATTGCACAGTAACAGTACCTTTCTGTA}\\
Core S117& \seqsplit{CAAAATTACAATAACGGATTCGCCATAAATCA}\\
Core S118& \seqsplit{TTGAATAGGAAGGGTTAGAACCTTTAAAAG}\\
Core S119& \seqsplit{TGAATAATCCAAGTTACAAAATCGTTTGAATT}\\
Core S120& \seqsplit{CAATATAATCATTTCAATTACCTAAAACAA}\\
Core S121& \seqsplit{GAAGGGCCGCTCACAATTCCACGCCTGGGG}\\
Core S122& \seqsplit{TTGTTATCGATCGGTGCGGGCCTCTTTCCGGC}\\
Core S123& \seqsplit{TAATCATGCTGGCGAAAGGGGGATCGGCCT}\\
Core S124& \seqsplit{AAGGCGATCCCGGGTACCGAGCTCCCAGTCGG}\\
Core S125& \seqsplit{AGAGGATCTAAGTTGGGTAACGCCGCATCTGC}\\
Core S126& \seqsplit{TGCCAAGCACGACGTTGTAAAACGAGTATTAG}\\
Core S127& \seqsplit{ACTTTACACATTTGAGGATTTAGAACGGCCAG}\\
Core S128& \seqsplit{AGATAATAAACAATTCGACAACTAAATAAA}\\
Core S129& \seqsplit{CACTAACACCCGAACGTTATTAATTACCATAT}\\
Core S130& \seqsplit{TTTGAGTGTTATCTAAAATATCAACACCGC}\\
Core S131& \seqsplit{TGAGGAAGAACATTATCATTTTGCTATACTTC}\\
Core S132& \seqsplit{GGTCAGTTCAGAAGGAGCGGAATAATTCAT}\\
Core S133& \seqsplit{CCACGCTGTGAGTGAGCTAACTCAGTGTGAAA}\\
Core S134& \seqsplit{CGCCTGGCCTCACTGCCCGCTTTGAATTCG}\\
Core S135& \seqsplit{GTGAGACGTCGTGCCAGCTGCATTTCGACTCT}\\
Core S136& \seqsplit{TTGGGCGCGCGCGGGGAGAGGCGGGCCATTAA}\\
Core S137& \seqsplit{AAATACCGATAGCCCTAAAACATCTTTGCGTA}\\
Core S138& \seqsplit{GCGAACTGAACGAACCACCAGCACGTCAAT}\\
Core S139& \seqsplit{ATATTTTTTGAGGCGGTCAGTATTTTTAGGAG}\\
Core S140& \seqsplit{CCTGAAAGAGTGCCACGCTGAGAGAAAGGAAT}\\
Core S141& \seqsplit{CATTCTGGAATCTAAAGCATCACAATATCT}\\
Core S142& \seqsplit{TGCCTAAGTTTGCCCCAGCAGG}\\
Core S143& \seqsplit{GAAACCTGGGCAACAGCTGATTGC}\\
Core S144& \seqsplit{CTGCAACCGTAAGAATACGTGG}\\
Core S145& \seqsplit{TTGCTTTCGGGATTTTGCTAAACGAACCCATGTACCG}\\
Core S146& \seqsplit{AAGGAGCCAGCGGAGTGAGAATAGACCCTCATTTTCAGG}\\
Core S147& \seqsplit{TCCAGTAAGTACTGGTAATAAGTGGAGGTTTAGTACC}\\
Core S148& \seqsplit{GAATGGAACCTTGAGTAACAGTGCTATAGCCCGGAATAG}\\
Core S149& \seqsplit{GGTCAGTGAGCGCAGTCTCTGAATATCACCGGAACCAGAGGTCAG}\\
Core S150& \seqsplit{TATTATTCCAGGTCAGACGATTGGACCCTCAGAGCCACCAACCA}\\
Core S151& \seqsplit{GACGCTGAGAGTGAATAACCTTGCTTTACATCGGGAGAAATTTGC}\\
Core S152& \seqsplit{ATTTTTGAAATTGTAAACGTTAATTAGCCAGCTTTCATCATCGCA}\\
\end{longtblr}

\clearpage

\begin{longtblr}[caption = {Staples for rectangular origami with poly-T overhangs}]
{
  colspec = {|XX[4]|},
  hlines,
} 
Core ST1& \seqsplit{TTTTTTTTTTTTTTTTTTTTATACCGATTTTCCAGACGTTAGTAACCAGTACAAACTAC}\\
Core ST2& \seqsplit{TTTTTTTTTTTTTTTTTTTTTTCACAAAATGCCCCCTGCCTATTGGATAAGTGCCGTCG}\\
Core ST3& \seqsplit{TTTTTTTTTTTTTTTTTTTTGGTTGAGGTGAAACATGAAAGTAAGGATTAGCGGGGT}\\
Core ST4& \seqsplit{TTTTTTTTTTTTTTTTTTTTTGTCGTCAGTTGCGCCGACAATATTCGGTC}\\
Core ST5& \seqsplit{TTTTTTTTTTTTTTTTTTTTTTCTGTATGAGGTGAATTTCTTAAGGCCGCTTTTGCGGGAAAGAA}\\
Core ST6& \seqsplit{TTTTTTTTTTTTTTTTTTTTACAGTTTCTTTAATTGTATCGGTTGCGAAAGA}\\
Core ST7& \seqsplit{TTTTTTTTTTTTTTTTTTTTAACTAAAAATCTCCAAAAAAAAGGCTACAGAGGCTTTCATTAAA}\\
Core ST8& \seqsplit{TTTTTTTTTTTTTTTTTTTTACGTTGAAGGAATTGCGAATAATATTGATGAT}\\
Core ST9& \seqsplit{TTTTTTTTTTTTTTTTTTTTACAGGAGTGCGTCATACATGGCTTATTTTTTC}\\
Core ST10& \seqsplit{TTTTTTTTTTTTTTTTTTTTACAGTTACAAATAAATCCTCATCTCCCTCA}\\
Core ST11& \seqsplit{TTTTTTTTTTTTTTTTTTTTGCTGAGGCAGCGATTATACCAAATCGCCTG}\\
Core ST12& \seqsplit{TTTTTTTTTTTTTTTTTTTTTTGACCCCCTTGCAGGGAGTTAAAACAGCTTG}\\
Core ST13& \seqsplit{TTTTTTTTTTTTTTTTTTTTAGAGGCAATCGTCACCCTCAGCATATCAGC}\\
Core ST14& \seqsplit{TTTTTTTTTTTTTTTTTTTTCAGCATCGCGAAGGCACCAACCTAGCAGACGG}\\
Core ST15& \seqsplit{TTTTTTTTTTTTTTTTTTTTTGCCACTAGAACGAGGGTAGCAACGGCTCCAA}\\
Core ST16& \seqsplit{TTTTTTTTTTTTTTTTTTTTGAAGTTTCGAGGACTAAAGACTTTTATTAGCG}\\
Core ST17& \seqsplit{TTTTTTTTTTTTTTTTTTTTTTTGCCATTCGGTCATAGCCCCCTTTCATGAG}\\
Core ST18& \seqsplit{TTTTTTTTTTTTTTTTTTTTCGGCATTTCTTTTCATAATCAAATTACCGT}\\
Core ST19& \seqsplit{TTTTTTTTTTTTTTTTTTTTCCTTTAGCCCACCACCGGAACCGCTAAAGCCA}\\
Core ST20& \seqsplit{TTTTTTTTTTTTTTTTTTTTGAGCCGCCAGTAGCGACAGAATAATTATTC}\\
Core ST21& \seqsplit{TTTTTTTTTTTTTTTTTTTTACCGTAATCACCCTCAGAACCGCCCCTTGATA}\\
Core ST22& \seqsplit{TTTTTTTTTTTTTTTTTTTTGAAACGTCCCCTCAGAGCCGCCAGACAGGA}\\
Core ST23& \seqsplit{TTTTTTTTTTTTTTTTTTTTATAAATTTTGCCCTGACGAGAATGGTTTAA}\\
Core ST24& \seqsplit{TTTTTTTTTTTTTTTTTTTTAATAAGGCGTGTCGAAATCCGCGAACTCATCT}\\
Core ST25& \seqsplit{TTTTTTTTTTTTTTTTTTTTCAAATCAATAGCCGGAACGAGGCAAACGAA}\\
Core ST26& \seqsplit{TTTTTTTTTTTTTTTTTTTTTCAATCATACAAGAACCGGATATTATACCAGT}\\
Core ST27& \seqsplit{TTTTTTTTTTTTTTTTTTTTTAATCTTGAAGGGAACCGAACTGAATACGTAA}\\
Core ST28& \seqsplit{TTTTTTTTTTTTTTTTTTTTGGCGCATAGACAGATGAACGGTGTCAGCGCCA}\\
Core ST29& \seqsplit{TTTTTTTTTTTTTTTTTTTTAAGACAAAAATTCATATGGTTTACACAGACCA}\\
Core ST30& \seqsplit{TTTTTTTTTTTTTTTTTTTTCAATAGAAAGGGCGACATTCAACTTTTCAT}\\
Core ST31& \seqsplit{TTTTTTTTTTTTTTTTTTTTACACCACGGGTAAATATTGACGGACAAGTTTG}\\
Core ST32& \seqsplit{TTTTTTTTTTTTTTTTTTTTATTAAAGCAACATATAAAAGAAGTAAGCAG}\\
Core ST33& \seqsplit{TTTTTTTTTTTTTTTTTTTTAAAGGTGGGTGAATTATCACCGTCATAGCAGC}\\
Core ST34& \seqsplit{TTTTTTTTTTTTTTTTTTTTAGTATGTTTGGGAATTAGAGCCAAAGGCCG}\\
Core ST35& \seqsplit{TTTTTTTTTTTTTTTTTTTTTTTCAACAACCCTCGTTTACCATTTGCAAA}\\
Core ST36& \seqsplit{TTTTTTTTTTTTTTTTTTTTACTATCATTTTAATCATTGTGAATATTCAGTG}\\
Core ST37& \seqsplit{TTTTTTTTTTTTTTTTTTTTCAAAAGGAAAGAACTGGCTCATTCATTACC}\\
Core ST38& \seqsplit{TTTTTTTTTTTTTTTTTTTTCAGGACGTCAACTAATGCAGATACTACTGCGG}\\
Core ST39& \seqsplit{TTTTTTTTTTTTTTTTTTTTACCACATTTGGGAAGAAAAATCTAATCAAGAG}\\
Core ST40& \seqsplit{TTTTTTTTTTTTTTTTTTTTTAGAAAGACTAACGGAACAACATTATAATAAG}\\
Core ST41& \seqsplit{TTTTTTTTTTTTTTTTTTTTAGCAAGAAAATTGAGTTAAGCCCAATTACAGG}\\
Core ST42& \seqsplit{TTTTTTTTTTTTTTTTTTTTCCCACAAGACAATGAAATAGCAATCACAAT}\\
Core ST43& \seqsplit{TTTTTTTTTTTTTTTTTTTTGGGTAATTGCCCTTTTTAAGAAAAACGCAAAG}\\
Core ST44& \seqsplit{TTTTTTTTTTTTTTTTTTTTATAGCCGTGAACACCCTGAACACAGTTACA}\\
Core ST45& \seqsplit{TTTTTTTTTTTTTTTTTTTTGAATTAACAACAAAGTTACCAGAAACATACAT}\\
Core ST46& \seqsplit{TTTTTTTTTTTTTTTTTTTTAACATAAACAATAATAACGGAATATTACGC}\\
Core ST47& \seqsplit{TTTTTTTTTTTTTTTTTTTTAGAAGTTGCGTTTTAATTCGAGAACAGGTC}\\
Core ST48& \seqsplit{TTTTTTTTTTTTTTTTTTTTCAAATATCTTGCCAGAGGGGGTAAAGAGCAAC}\\
Core ST49& \seqsplit{TTTTTTTTTTTTTTTTTTTTTGTTTAGAATTAAGAGGAAGCCCGGCTCCTTT}\\
Core ST50& \seqsplit{TTTTTTTTTTTTTTTTTTTTTCAAAAAGCTGGATAGCGTCCAAATAACGC}\\
Core ST51& \seqsplit{TTTTTTTTTTTTTTTTTTTTAATCGTCAAGTCAGAAGCAAAGCGTGGCTTAG}\\
Core ST52& \seqsplit{TTTTTTTTTTTTTTTTTTTTACTATTATTAAATATTCATTGAATTTAGGAAT}\\
Core ST53& \seqsplit{TTTTTTTTTTTTTTTTTTTTAATGCTTTCAAAAATCAGGTCTGTAGCTCA}\\
Core ST54& \seqsplit{TTTTTTTTTTTTTTTTTTTTACCATAAATAAACAGTTCAGAAAACAGCTACA}\\
Core ST55& \seqsplit{TTTTTTTTTTTTTTTTTTTTATTTTATCGTTGCTATTTTGCACCCGAGAATG}\\
Core ST56& \seqsplit{TTTTTTTTTTTTTTTTTTTTCAAGATTACTGAATCTTACCAACGAGATAA}\\
Core ST57& \seqsplit{TTTTTTTTTTTTTTTTTTTTGCGTCTTTCGGGAGGTTTTGAAGCTAGAAACC}\\
Core ST58& \seqsplit{TTTTTTTTTTTTTTTTTTTTCCGACTTGCCAGAGCCTAATTTGCAAGTCAGA}\\
Core ST59& \seqsplit{TTTTTTTTTTTTTTTTTTTTAAATAAACGCGAGGCGTTTTAGTATCATTC}\\
Core ST60& \seqsplit{TTTTTTTTTTTTTTTTTTTTTCTAAGAACAGCCATATTATTTATAGACGGGA}\\
Core ST61& \seqsplit{TTTTTTTTTTTTTTTTTTTTAAATAAGAAGATATAGAAGGCTTAACCGCACT}\\
Core ST62& \seqsplit{TTTTTTTTTTTTTTTTTTTTAGCAAATCAACGATTTTTTGTTTAGAGAAT}\\
Core ST63& \seqsplit{TTTTTTTTTTTTTTTTTTTTAGGATTATAGCTATATTTTCATCTACTAAT}\\
Core ST64& \seqsplit{TTTTTTTTTTTTTTTTTTTTAACCTGTTGAGAGTACCTTTAATTAAAGACTT}\\
Core ST65& \seqsplit{TTTTTTTTTTTTTTTTTTTTTGATAAGAAGATACATTTCGCAAAAATCATAC}\\
Core ST66& \seqsplit{TTTTTTTTTTTTTTTTTTTTTGACCATTGGTCATTTTTGCGGAGATTGCA}\\
Core ST67& \seqsplit{TTTTTTTTTTTTTTTTTTTTAGCTTAATTCTGCGAACGAGTAGAATTAAGCA}\\
Core ST68& \seqsplit{TTTTTTTTTTTTTTTTTTTTTTCCCAATTGCTGAATATAATGCTTTACCCTG}\\
Core ST69& \seqsplit{TTTTTTTTTTTTTTTTTTTTACATGTTGTTTCATTCCATATAAATCGGTT}\\
Core ST70& \seqsplit{TTTTTTTTTTTTTTTTTTTTGTCTGGAATTAAATATGCAACTAAAAATAATA}\\
Core ST71& \seqsplit{TTTTTTTTTTTTTTTTTTTTTCCCATCCAAGTCCTGAACAAGAAAGTACGGT}\\
Core ST72& \seqsplit{TTTTTTTTTTTTTTTTTTTTCAATAGATTAATTTACGAGCATGCTTAAAT}\\
Core ST73& \seqsplit{TTTTTTTTTTTTTTTTTTTTAATCAATATGCAGAACGCGCCTGTAGTATCAT}\\
Core ST74& \seqsplit{TTTTTTTTTTTTTTTTTTTTTCAGCTAAATCGGCTGTCTTTCCTCGAACCTC}\\
Core ST75& \seqsplit{TTTTTTTTTTTTTTTTTTTTCAAGAACGACGACGACAATAAATATAAAGC}\\
Core ST76& \seqsplit{TTTTTTTTTTTTTTTTTTTTTCTGTCCAGGGTATTAAACCAAGTTCCGGTAT}\\
Core ST77& \seqsplit{TTTTTTTTTTTTTTTTTTTTCATCGAGAAGTACCGACAAAAGGTTTGAGAAT}\\
Core ST78& \seqsplit{TTTTTTTTTTTTTTTTTTTTGAATATAAACAAGCAAGCCGTTTAATAGCA}\\
Core ST79& \seqsplit{TTTTTTTTTTTTTTTTTTTTAGTAGTAGGTGAGAAAGGCCGGACCGTTCT}\\
Core ST80& \seqsplit{TTTTTTTTTTTTTTTTTTTTTTCAAAAGGCATTAACATCCAATATGGTCAAT}\\
Core ST81& \seqsplit{TTTTTTTTTTTTTTTTTTTTAGGCAAGGGCCTGAGTAATGTGTAGGGTAGCT}\\
Core ST82& \seqsplit{TTTTTTTTTTTTTTTTTTTTAATGCAATCAAAGAATTAGCAAATTTAGTT}\\
Core ST83& \seqsplit{TTTTTTTTTTTTTTTTTTTTATAAAGCCTTTTAGAACCCTCATATATCAGGT}\\
Core ST84& \seqsplit{TTTTTTTTTTTTTTTTTTTTATAAAAATTCAGAGCATAAAGCTAACAGTTGA}\\
Core ST85& \seqsplit{TTTTTTTTTTTTTTTTTTTTGTACCAAGAAGCCTTTATTTCACAAGAGAA}\\
Core ST86& \seqsplit{TTTTTTTTTTTTTTTTTTTTTTGCGGGAAAACATTATGACCCTGCGGAATCA}\\
Core ST87& \seqsplit{TTTTTTTTTTTTTTTTTTTTTAATTACTAAATAAGAATAAACACTAATACTT}\\
Core ST88& \seqsplit{TTTTTTTTTTTTTTTTTTTTAAGGCGTTAGAAAAAGCCTGTTTTTATCAA}\\
Core ST89& \seqsplit{TTTTTTTTTTTTTTTTTTTTATGCGTTAGAAATACCGACCGTGTTAGATTAA}\\
Core ST90& \seqsplit{TTTTTTTTTTTTTTTTTTTTAATGGTTTTACAAATTCTTACCAGCAACATGT}\\
Core ST91& \seqsplit{TTTTTTTTTTTTTTTTTTTTCAACGCTATTTCATCTTCTGACATTTATCA}\\
Core ST92& \seqsplit{TTTTTTTTTTTTTTTTTTTTTTTAGTTACAACAGTAGGGCTTAAAAAGTAAT}\\
Core ST93& \seqsplit{TTTTTTTTTTTTTTTTTTTTCGCCATATGCGAGAAAACTTTTTCCTTTTTAA}\\
Core ST94& \seqsplit{TTTTTTTTTTTTTTTTTTTTCAAAGAACTTAACAACGCCAACAAATAAGA}\\
Core ST95& \seqsplit{TTTTTTTTTTTTTTTTTTTTAGCTGATCATTAAATTTTTGTTCATCAAAA}\\
Core ST96& \seqsplit{TTTTTTTTTTTTTTTTTTTTAAAATTCGAAATTAATGCCGGAGAGGTAAAGA}\\
Core ST97& \seqsplit{TTTTTTTTTTTTTTTTTTTTAATATTTAGAGATCTACAAAGGCTATTTTA}\\
Core ST98& \seqsplit{TTTTTTTTTTTTTTTTTTTTCATTGCCTAAAACAGGAAGATTGTGAGTAACA}\\
Core ST99& \seqsplit{TTTTTTTTTTTTTTTTTTTTAAGCCCCAGAGAGTCTGGAGCAAAACGCAAGG}\\
Core ST100& \seqsplit{TTTTTTTTTTTTTTTTTTTTTCGATGATGTACCCCGGTTGATAAACGGCGGATTGACGCGCATC}\\
Core ST101& \seqsplit{TTTTTTTTTTTTTTTTTTTTCAATCATAACGGTAATCGTAAAACCCTTAGAA}\\
Core ST102& \seqsplit{TTTTTTTTTTTTTTTTTTTTTCCTTGAACTATTAATTAATTTTCTAGCATGT}\\
Core ST103& \seqsplit{TTTTTTTTTTTTTTTTTTTTAATCGTCGAACATAGCGATAGCTGATAAAT}\\
Core ST104& \seqsplit{TTTTTTTTTTTTTTTTTTTTATATATGTGAAGAGTCAATAGTGACTAAATTT}\\
Core ST105& \seqsplit{TTTTTTTTTTTTTTTTTTTTAAATCATTAATGGAAACAGTACTGATTGCT}\\
Core ST106& \seqsplit{TTTTTTTTTTTTTTTTTTTTACCTTTTTAGGTCTGAGAGACTACAAATATAT}\\
Core ST107& \seqsplit{TTTTTTTTTTTTTTTTTTTTCCTCCGGCACATTTAACAATTTCACGCAGAGGCGAATTATTCCTG}\\
Core ST108& \seqsplit{TTTTTTTTTTTTTTTTTTTTAATTAATTTTAGGTTGGGTTATACGCAAGA}\\
Core ST109& \seqsplit{TTTTTTTTTTTTTTTTTTTTATAATTCTGGTGCCGGAAACCAACTGTTGG}\\
Core ST110& \seqsplit{TTTTTTTTTTTTTTTTTTTTACCGCTTCGCGTCTGGCCTTCCTGATTTTGTT}\\
Core ST111& \seqsplit{TTTTTTTTTTTTTTTTTTTTCAGGAAGAACATTAAATGTGAGCATAAGCA}\\
Core ST112& \seqsplit{TTTTTTTTTTTTTTTTTTTTACCCGTCGGGGGACGACGACAGTATGTGCTGC}\\
Core ST113& \seqsplit{TTTTTTTTTTTTTTTTTTTTCAGTTTGAGATTCTCCGTGGGAACAATCAGAA}\\
Core ST114& \seqsplit{TTTTTTTTTTTTTTTTTTTTGTAGATGGCGTAATGGGATAGGTCAACGTCAG}\\
Core ST115& \seqsplit{TTTTTTTTTTTTTTTTTTTTATGAATATGTAGATTTTCAGGTTTACGTTGGT}\\
Core ST116& \seqsplit{TTTTTTTTTTTTTTTTTTTTGAAATTGCACAGTAACAGTACCTTTCTGTA}\\
Core ST117& \seqsplit{TTTTTTTTTTTTTTTTTTTTCAAAATTACAATAACGGATTCGCCATAAATCA}\\
Core ST118& \seqsplit{TTTTTTTTTTTTTTTTTTTTTTGAATAGGAAGGGTTAGAACCTTTAAAAG}\\
Core ST119& \seqsplit{TTTTTTTTTTTTTTTTTTTTTGAATAATCCAAGTTACAAAATCGTTTGAATT}\\
Core ST120& \seqsplit{TTTTTTTTTTTTTTTTTTTTCAATATAATCATTTCAATTACCTAAAACAA}\\
Core ST121& \seqsplit{TTTTTTTTTTTTTTTTTTTTGAAGGGCCGCTCACAATTCCACGCCTGGGG}\\
Core ST122& \seqsplit{TTTTTTTTTTTTTTTTTTTTTTGTTATCGATCGGTGCGGGCCTCTTTCCGGC}\\
Core ST123& \seqsplit{TTTTTTTTTTTTTTTTTTTTTAATCATGCTGGCGAAAGGGGGATCGGCCT}\\
Core ST124& \seqsplit{TTTTTTTTTTTTTTTTTTTTAAGGCGATCCCGGGTACCGAGCTCCCAGTCGG}\\
Core ST125& \seqsplit{TTTTTTTTTTTTTTTTTTTTAGAGGATCTAAGTTGGGTAACGCCGCATCTGC}\\
Core ST126& \seqsplit{TTTTTTTTTTTTTTTTTTTTTGCCAAGCACGACGTTGTAAAACGAGTATTAG}\\
Core ST127& \seqsplit{TTTTTTTTTTTTTTTTTTTTACTTTACACATTTGAGGATTTAGAACGGCCAG}\\
Core ST128& \seqsplit{TTTTTTTTTTTTTTTTTTTTAGATAATAAACAATTCGACAACTAAATAAA}\\
Core ST129& \seqsplit{TTTTTTTTTTTTTTTTTTTTCACTAACACCCGAACGTTATTAATTACCATAT}\\
Core ST130& \seqsplit{TTTTTTTTTTTTTTTTTTTTTTTGAGTGTTATCTAAAATATCAACACCGC}\\
Core ST131& \seqsplit{TTTTTTTTTTTTTTTTTTTTTGAGGAAGAACATTATCATTTTGCTATACTTC}\\
Core ST132& \seqsplit{TTTTTTTTTTTTTTTTTTTTGGTCAGTTCAGAAGGAGCGGAATAATTCAT}\\
Core ST133& \seqsplit{TTTTTTTTTTTTTTTTTTTTCCACGCTGTGAGTGAGCTAACTCAGTGTGAAA}\\
Core ST134& \seqsplit{TTTTTTTTTTTTTTTTTTTTCGCCTGGCCTCACTGCCCGCTTTGAATTCG}\\
Core ST135& \seqsplit{TTTTTTTTTTTTTTTTTTTTGTGAGACGTCGTGCCAGCTGCATTTCGACTCT}\\
Core ST136& \seqsplit{TTTTTTTTTTTTTTTTTTTTTTGGGCGCGCGCGGGGAGAGGCGGGCCATTAA}\\
Core ST137& \seqsplit{TTTTTTTTTTTTTTTTTTTTAAATACCGATAGCCCTAAAACATCTTTGCGTA}\\
Core ST138& \seqsplit{TTTTTTTTTTTTTTTTTTTTGCGAACTGAACGAACCACCAGCACGTCAAT}\\
Core ST139& \seqsplit{TTTTTTTTTTTTTTTTTTTTATATTTTTTGAGGCGGTCAGTATTTTTAGGAG}\\
Core ST140& \seqsplit{TTTTTTTTTTTTTTTTTTTTCCTGAAAGAGTGCCACGCTGAGAGAAAGGAAT}\\
Core ST141& \seqsplit{TTTTTTTTTTTTTTTTTTTTCATTCTGGAATCTAAAGCATCACAATATCT}\\
Core ST142& \seqsplit{TTTTTTTTTTTTTTTTTTTTTGCCTAAGTTTGCCCCAGCAGG}\\
Core ST143& \seqsplit{TTTTTTTTTTTTTTTTTTTTGAAACCTGGGCAACAGCTGATTGC}\\
Core ST144& \seqsplit{TTTTTTTTTTTTTTTTTTTTCTGCAACCGTAAGAATACGTGG}\\
Core ST145& \seqsplit{TTTTTTTTTTTTTTTTTTTTTTGCTTTCGGGATTTTGCTAAACGAACCCATGTACCG}\\
Core ST146& \seqsplit{TTTTTTTTTTTTTTTTTTTTAAGGAGCCAGCGGAGTGAGAATAGACCCTCATTTTCAGG}\\
Core ST147& \seqsplit{TTTTTTTTTTTTTTTTTTTTTCCAGTAAGTACTGGTAATAAGTGGAGGTTTAGTACC}\\
Core ST148& \seqsplit{TTTTTTTTTTTTTTTTTTTTGAATGGAACCTTGAGTAACAGTGCTATAGCCCGGAATAG}\\
Core ST149& \seqsplit{TTTTTTTTTTTTTTTTTTTTGGTCAGTGAGCGCAGTCTCTGAATATCACCGGAACCAGAGGTCAG}\\
Core ST150& \seqsplit{TTTTTTTTTTTTTTTTTTTTTATTATTCCAGGTCAGACGATTGGACCCTCAGAGCCACCAACCA}\\
Core ST151& \seqsplit{TTTTTTTTTTTTTTTTTTTTGACGCTGAGAGTGAATAACCTTGCTTTACATCGGGAGAAATTTGC}\\
Core ST152& \seqsplit{TTTTTTTTTTTTTTTTTTTTATTTTTGAAATTGTAAACGTTAATTAGCCAGCTTTCATCATCGCA}\\
\end{longtblr}

\clearpage

\begin{longtblr}[caption = {Remaining structural staples for rectangular origami}]
{
  colspec = {|XX[4]|},
  hlines,
} 
core1&AAATCAAAAGAATAGCCAAGCGGT\\
core2&GTGTTGTTCCAGTTTGCCTTCAC\\
core3&ATTAAAGAACGTGGACTTTTCACCA\\
core4&CTTGCTGGTAATATCCTTAATGC\\
core5&CCAGCCATTGCAACAGGCACAGACA\\
core6&AATACCTACATTTTGACCTTCTGA\\
core7&AATGGATTATTTACATAAAGGGA\\
core8&ATTGTTTGGATGGAACAAAGAAACCACGGCAAATCAACAGTTG\\
core9&CGGGTAAACCAACTTTGAAAGAGGGCTGGCTGACCTTC\\
core10&GTAACCGTAGGGTTTTCCCAGTCTTGCATGCCTGCAGG\\
core11&TACACTAAAACCCTGCTCCATGTTACTCGTAACAAAGCTGCTC\\
core12&GGAAACCGAGGAAACGAACAGGGAAGCGCATTCCCAATCC\\
core13&TACCTTATGCGATTTTATTACGAGGCATAGTATAGTAAAA\\
core14&CGTTAATAAAACGAATTCATCAGTTGAGATCCCCCTCA\\
core15&TAGCTATCTTACCGAAGAGCGCTAATATCAGAGCTAACGA\\
core16&ACGTAAAACAGCGTATTAAATCCTTTGACTAATAGATTAGAGC\\
core17&CTCCAGCCAGCTTCGCTATTACGCCAGGTCATAGCTGTTTCCT\\
core18&ACTGTAGCGCGCGATTGAGGGAGGGAAGAATAAGTTTATTTTG\\
core19&ATGAAACCATCGACCGACTTGAGCCATTAGCAAACGTAGAAAAT\\
\end{longtblr}

\clearpage

\begin{longtblr}[caption = {Handle strands on the edges of the rectangular origami and the lock strand to connect the handle strands}]
{
  colspec = {|XX[4]|},
  hlines,
} 
Handle top1& \seqsplit{GAAAAACCGTCTATCAACTCAAACTATCGGCTTTTTTCGAGTTCA}\\
Handle top2& \seqsplit{CCAGCAGCAAATGAAACCAACAGAGATAGAACCGCTCAATCGTCTGATTTTTTCGAGTTCA}\\
Handle top3& \seqsplit{GAAGATAAAACAGAGGGAATGGCTATTAGTCTAGAACAATATTACCGTTTTTTCGAGTTCA}\\
Handle top4& \seqsplit{AATGAATCGGCCAACCAGGGTGGTTTTTCTCCAACGTCAAAGGGCTTTTTTCGAGTTCA}\\
Handle top5& \seqsplit{CATTAATTGCGTTGCGCCTGAGAGAGTTGCAGCCGAGATAGGGTTGATTTTTTCGAGTTCA}\\
Handle top6& \seqsplit{GAACAAGAGTCCACTTTTTTTCGAGTTCA}\\
Handle top7& \seqsplit{GGCAAAATCCCTTATTTTTTTCGAGTTCA}\\
Handle top8& \seqsplit{AAAAACGCTCATGGATTTTTTCGAGTTCA}\\
Handle bottom1& \seqsplit{CTTGTTGGTTTTTTAACGCCTGTAGCATTCTAAAGTTT}\\
Handle bottom2& \seqsplit{CTTGTTGGTTTTTTTAACACTGAGTTTCGTCAATGAATT}\\
Handle bottom3& \seqsplit{CTTGTTGGTTTTTTGATAGCAAGCCCAATAGAACTTTCA}\\
Handle bottom4& \seqsplit{CTTGTTGGTTTTTTCACCCTCAGAGCCACCAAAGGAAC}\\
Handle bottom5& \seqsplit{CTTGTTGGTTTTTTGTGTATCACCGTACTCATTTAACGG}\\
Handle bottom6& \seqsplit{CTTGTTGGTTTTTTAGAGGGTTGATATAAGCCGTATAA}\\
Handle bottom7& \seqsplit{CTTGTTGGTTTTTTTTTGCTCAGTACCAGGCTCGGAACC}\\
Handle bottom8& \seqsplit{CTTGTTGGTTTTTTGCCACCCTCAGAACCGCCACCCTCAGAACCGC}\\
lock&CCAACAAGTGAACTCG\\
Biotinylated bridging strand& \seqsplit{AAAAAAAAAAAAAAAAAAAA/3Bio/}\\
\end{longtblr}

\bibliographystyle{old-mujstyl}
\bibliography{newrefs}